\documentclass[useAMS,usenatbib]{mn2e}
\PassOptionsToPackage{square,comma,numbers,sort&compress,super}{natbib}
\usepackage[colorlinks=true,urlcolor=blue,citecolor=red,linkcolor=red,bookmarks=true]{hyperref}

\usepackage{graphicx}

\usepackage[font=small]{caption}

\title[Rotation curves in $\Lambda$-CDM]{Rotation curves of galaxies and the stellar mass-to-light ratio }

\author[Haghi et al.]
{Hosein Haghi$^{1}$\thanks{
E-mail:  \mbox{haghi@iasbs.ac.ir} (HH)
 },  Aziz Khodadadi$^{1}$, Amir Ghari$^{1}$, Akram Hasani Zonoozi$^{1}$, \\\\
\newauthor
Pavel Kroupa$^{2, 3}$\\
$^{1}$Department of Physics, Institute for Advanced Studies in Basic Sciences (IASBS), P.O. Box 11365-9161, Zanjan, Iran\\
$^{2}$Helmholtz-Institut f\"ur Strahlen-und Kernphysik (HISKP), Universit\"at Bonn, Nussallee 14-16, D-53115 Bonn, Germany\\
$^{3}$Charles University in Prague, Faculty of Mathematics and Physics, Astronomical Institute, V Hole\v{s}ovi\v{c}k\'ach 2, CZ-180 00 \\
Praha 8, Czech Republic\\}

\begin{document}

\date{Accepted \ldots. Received \ldots; in original form \ldots}

\pagerange{\pageref{firstpage}--\pageref{lastpage}} \pubyear{2013}

\maketitle

\label{firstpage}

\maketitle

\begin{abstract}

Mass models of a sample of 171 low- and high-surface brightness galaxies are presented in the context of the cold dark matter (CDM) theory using the NFW dark matter halo density distribution to extract a new concentration-viral mass relation ($c-M_{vir}$). The  rotation curves (RCs) are calculated from the total baryonic matter based on the 3.6 $\mu m $-band surface photometry, the observed distribution of neutral hydrogen, and the dark halo, in which the three adjustable parameters are the stellar mass-to-light ratio, halo concentration and virial mass. Although accounting for a NFW dark halo profile can explain rotation curve observations, the implied $c-M_{vir}$ relation from RC analysis strongly disagrees with that resulting from different cosmological simulations. Also, the $M/L-$color correlation of the studied galaxies is inconsistent with that expected from stellar population synthesis models with different stellar initial mass functions. Moreover, we show that the best-fitting stellar $M/L-$ ratios of 51 galaxies (30\% of our sample) have unphysically negative values in the framework of the $\Lambda $CDM theory.  This can be interpreted as a serious crisis for this theory. This suggests either that the commonly used NFW halo profile, which is a natural result of $\Lambda $CDM cosmological structure formation, is not an appropriate profile for the dark halos of galaxies, or, new dark matter physics or alternative gravity models are needed to explain the rotational velocities of disk galaxies.

\end{abstract}


\section{Introduction}

According to Newtonian gravity, the rotational velocity falls with distance from the center of a galaxy (the so-called Keplerian fall-off), while the observed data for galaxies usually show an asymptotically flat rotation curve out to the furthest observationally accessible data points. The flattening of the rotational velocity of material for a large sample of spiral galaxies is strong empirical  evidence for the discrepancy between the visible mass and the dynamical mass of galaxies \citep{rub70, rub78, bos78}.  Rotation curves of spiral galaxies have been studied for several decades now, and provide a valuable body of data to determine the radial dependency of the gravitational forces on galactic scales.

The generally accepted explanation of the galaxy mass discrepancy problem is the  cold dark matter (CDM) model in which the visible disk of a spiral galaxy (in the form of gas and stars) might be surrounded by a more massive and extensive halo of unseen cold dark matter (CDM; \citealt{beg91,per96,chem11}) made of non-baryonic at most weakly interacting particles which are not described by the standard model of physics (e.g. \citealt{bert05}) and which dominates the gravitational field in the outer parts \citep{korknap87}.  In the $\Lambda$-CDM framework, the dark halos merge in a hirarchical fashion into larger halos, building up over time the present-day galaxies.

Dark matter (DM) had also been inferred to contribute significantly on larger scales in the Universe from the large velocity dispersions observed in galaxy clusters by \cite{Zwic} in the 1930s, gravitational lensing of background objects by galaxy clusters such as the Bullet Cluster, the temperature distribution of hot gas in galaxies and clusters of galaxies, and more recently by the pattern of anisotropies in the cosmic microwave background.  Despite more than 40 years of extensive searches for dark matter particles, no well-agreed candidate particles of DM have yet been directly detected. The lack of evidence for dynamical friction due to dark matter halos also challenges the existence of dark matter particles (Kroupa 2015; Oehm, Thies \& Kroupa 2017).

Theoretically, several suggestions are proposed in the literature for the density profile models of the DM distribution. One of the most
important commonly used models is the NFW  dark matter halo model \citep{nav96},  derived from  $\Lambda$-CDM cosmological simulations of structure formation using collisionless DM particles. This work suggests that equilibrium DM haloes, produced through hierarchical clustering, are well approximated by a universal, two-parameter density profile.

CDM model \citep{nav96,moo99,nav04} predictions for a sample of spiral galaxies with accurately measured RCs concluded that the CDM
hypothesis  fails to reproduce observed RCs (see, e.g., \citealt{deb01, deb02, gen04, gen05, gen07b, gen07c, mcg07, zon10, wu15}).
Deriving the rotation curves of 19 galaxies of the THINGS sample, \cite{deb08} found that most of these galaxies preferred the observationally motivated core-dominated isothermal halo (i.e., the shallower central region) over the cuspy NFW haloes, the so-called core-cusp controversy. \cite{zon10} constructed RCs of a large sample of 48 galaxies from the distribution of their detectable matter through a set of different gravity models. While the different models reproduce the observed data with reasonable detail, on a deeper examination, they found significant disparities in their predictions of stellar mass -to-light ($M_*/L$) ratios in the framework of the CDM theory using the NFW profile for galaxy halos. They also showed that the stellar population synthesis (SPS) analysis and the color $M_*/L$ correlation predicted therein through various initial mass functions (IMF) could differentiate between the gravity models.

In this paper we aim at fitting the RCs of a large sample of 171 galaxies from the new \emph{Spitzer} Photometry and Accurate Rotation Curves (SPARC) data set using the NFW model for dark matter halo profile. The paper is organized as follow: the rotation curve data of a sample of galaxies are described in Section \ref{data}.  In Section \ref{fit} we produce the rotation curve fits to the used sample of galaxies using the NFW halo profile. This is followed by a presentation of the results of fits in Section \ref{result}. Conclusions are contained in Section \ref{conclusion}.

\section{The data}\label{data}
We used RCs from the SPARC dataset which is compiled from the literature by \citet{Lelli16}. The sample includes a collection of 171 galaxies, spans a wide range of K-band luminocities from $ 3\times 10^7$ to  $ 3\times 10^{11} L_{\odot,K}$  and morphological types, from gas-dominated low-surface brightness (LSB) galaxies (e.g., DDO 154 and IC 2574) to high surface brightness (HSB) galaxies with a massive stellar component and a low gas content with well-extended rotation curves (e.g., NGC 5033). This diversity is useful for studying dynamical properties of spiral galaxies. The properties of a subset of this sample with the negative best-fitting stellar $M_*/L$ ratios are listed in Table \ref{t1} and the corresponding RCs are shown in Fig. \ref{f2} (see Section \ref{result} for more details).  The details of the data sample are explained in the SPARC main paper \citep{Lelli16}.

\section{Fitting the Rotation Curves}\label{fit}

In fitting RCs we follow the procedure outlined by \cite{beg91}. The rotation curve includes contributions from the stellar and gaseous disk, stellar bulge if any, and the dark matter halo. The stellar components used in this study are derived from 3.6 $\mu m $-band surface brightness profiles (which is a good proxy for the emission of the stellar disc), converted into mass using the $M_*/L_{[3.6]}$ ratio (as a free parameter) in that particular band (See \cite{Lelli16} for more details).   This $M_*/L_{[3.6]}$  ratio is assumed to be constant with radius, throughout the galaxy, though this is not strictly the case, because of the color gradient in spiral galaxies. We leave the amplitude of the disk contribution to the circular velocity (i.e. the disk mass) as a free parameter to be derived by fitting the RC.  The atomic hydrogen gas contribution is derived from the HI maps corrected for the primordial helium contribution by scaling the HI mass by a factor of 1.4. We assume that the dark matter halo has a density distribution given by a NFW profile,

\begin{equation}\label{eq:nfwprofile}
 \rho_{NFW}(r) = \frac {\rho_0}{r/r_s (1+r/r_s)^2},
\end{equation}
where $\rho_0$ and $r_s$ are scaling parameters that characterize a given halo. By integrating Eq.~\ref{eq:nfwprofile}, the total mass inside the radius $r$ is given by
\begin{equation}
 M_{NFW}(r) = 4 \pi \rho_0 r_s ^3 \left[\ln(1+r/r_s)-\frac{r/r_s}{1+r/r_s}\right].
\end{equation}

The NFW density profile can be equivalently identified by just two parameters, the virial mass of the halo, $M_{vir}$, and the concentration, $c = r_{vir}/r_s$, which relates the inner and virial parameters as
\begin{eqnarray}
 \rho_0 &=& \frac{\rho_{crit} \Delta_{vir}}{3} \frac{c^3}{\ln(1+c) - c/(1+c)},\\
 r_s &=& \frac{1}{c} \left(\frac{3 M_{vir}}{4\pi \Delta_{vir} \rho_{crit}}  \right)^{1/3},
\end{eqnarray}
where $\Delta_{vir}$ is the virial overdensity criterion, which is a function of cosmology and redshift and
varies from 100 to 200,  $\rho_{crit} = 3 H^2(z)/8 \pi G$ is the critical density and  $r_{vir}$ is the virial radius of a sphere that encloses an average
density of $\Delta_{vir}$ $\times$ $\rho_{crit}$  within the virialized region. It is conventional to describe a DM halo as a spherical region with density of about $200\rho_{crit}$ of the universe. Hence the virial mass of this spherical over-dense region is given by:
\begin{equation}
M_{vir} = \frac {4}{3} \pi \Delta_{vir} \rho_{crit} r_{vir}^3 ,~~ with~~ \Delta_{vir}=200.
\end{equation}

The potential to which the NFW density profile corresponds is given by
\begin{equation}
\Phi(r) = - \frac {G M_{vir}}{r}  \frac{\ln(1+r/r_s)}{\ln(1+c)- c/(1+c)}.
\end{equation}
Simulations of structure formation have shown that there exists a (redshift-dependent) correlation between halo concentration, $c$, and the virial mass, $M_{vir}$, such that more massive halos are less concentrated \citep{bull01, wechsler02, neto07, klypin11}. The concentration-mass relation for virialised halos can be approximated as
\begin{equation}\label{eq:c-mvir}
\log_{10}\left( c\right) = 1.025-0.097\,\log_{10}\left(\frac{M_{vir}}{10^{12}h^{-1}}\right).
\end{equation}
Therefore, the NFW halo profile can be rewritten, at a given redshift, in terms of a single parameter, $M_{vir}$. Eq. 7 is valid for redshift $z=0$.

The model RCs can be presented as a quadratic sum of the circular velocities of the various components:
\begin{equation}
v_{rot}^2=v_{d}^2(M_*/L_{3.6})_d + v_{b}^2( M_*/L_{3.6})_b + v_{g}^2 + v_{h}^2,
\end{equation}
where,  $v_d$,  $v_b$, $v_{g}$, and $v_h$ is the contribution of the stellar disk, bulge gas and DM halo to the rotation curve, respectively. $(M_*/L)_d$ and $(M_*/L)_b$ are the  3.6 micron band stellar mass-to-light ratios of the disk and the bulge. In this work we adopt $(M_*/L)_b= 1.4 (M_*/L)_d$ as suggested by SPS models (Schombert \& McGaugh 2014). \cite{McGaugh16} shows that this is a very good approximation.

We quantify the  goodness-of-fit by computing the reduced $\chi^2$. Fitting of the calculated rotation curves to the observed data points is achieved by finding the stellar $M_*/L_{[3.6]}$ ratio and halo parameters in parameter space, by minimizing the reduced least-squares value,
\begin{equation}
 \chi^2_{\nu}=\frac{1}{(N-P)}\sum_{i=1}^{N}\frac{(v^{i}_{t}-v_{o}^{i})^2}{\sigma_i^2}\label{chi},
\end{equation}
\noindent where $\sigma_i$ is the observational uncertainty in the rotation speeds and $P$ is the number of degrees of freedom. $N$ is the number of observed velocity values along the radial direction in a galaxy. The $M_*/L_{[3.6]}$ ratio of the stellar component, halo concentration $c$,  and the halo virial mass $M_{vir}$ are free parameters (i.e., P=3), where allowed to $M_*/L_{[3.6]}$ vary in a physically relevant range.

\section{Results} \label{result}
In this section we present the results  of the RC fits (shown in Figs. \ref{f2} and summarized in Table \ref{t1}) by comparing the calculated circular velocity curves in the $\Lambda$CDM framework with the observed RCs using the least square algorithm (Eq. \ref{chi}).

\subsection{Negative stellar mass-to-light ratios }

\begin{figure}
\includegraphics[width=85mm]{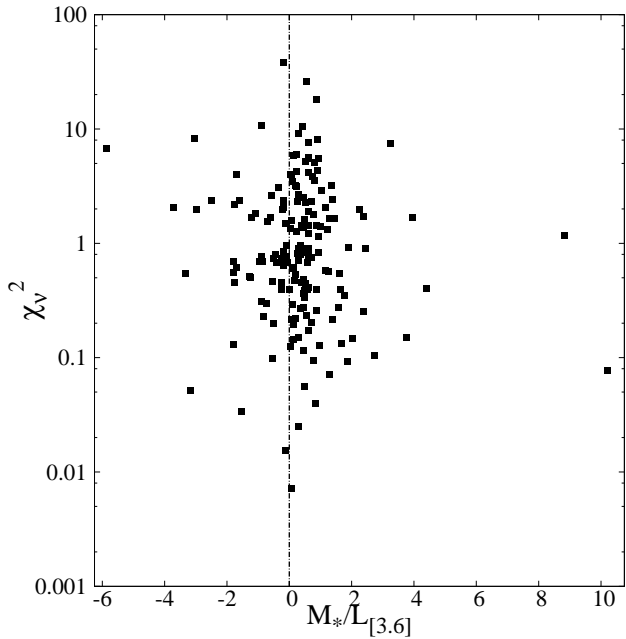}
\caption{$\chi^2_{\nu}$ vs. $M_*/L_{[3.6]}$. For 41 galaxies in the sample, $\chi^2_{\nu}$ is larger than 3. The implied stellar $M_*/L_{[3.6]}$ ratio in 51 out of the 171 galaxies of our sample have unphysically negative values as listed in Table \ref{t1}
} \label{chi-gam}
\end{figure}

The NFW dark matter halo fits are made with the $M_*/L_{[3.6]}$ ratio, $M_{vir}$, and $c$ as three fitting parameters.  According to the $\chi^2_{\nu}$ values in Table \ref{t1}, despite the flexibility provided by three fitting parameters, the matching of the observed RCs is largely good with $\chi^2_{\nu}\approx1$. Figure \ref{chi-gam} shows $\chi^2_{\nu}$ versus $M_*/L_{[3.6]}$.  In particular, discrepancies between the theoretically constructed best-fitting RCs and the observed ones are seen for 31 galaxies in the sample (with $\chi^2_{\nu}\geq 3$).  Moreover, the implied stellar $M_*/L_{[3.6]}$ ratio in 51 out of the 171 galaxies of our sample have unphysically negative values as listed in Table \ref{t1}. It should be noted that the average value of the best fitted stellar $M_*/L_{[3.6]}$ ratios of our sample is 0.54 which is close to the adopted constant $M_*/L_{[3.6]}$ value of 0.5 presented in \citet{Schombert2012}.  

In  Fig. \ref{f2} we show the results of the three-parameter NFW fits of theoretically constructed RCs to the observations of these 51 galaxies where the RCs of the individual components are also shown. In each panel, the measured rotational velocities are indicated as black points with error-bars and the best-fitting RCs are shown by black lines. The stellar disk and gas and dark halo contributions are represented as the red, green and blue lines, respectively.

In Table \ref{t1}, we tabulate the key results of the best fitted RCs, that is, $\chi^2_{\nu}$, $M_{vir}$, $c$ and the implied $M_*/L_{[3.6]}$ ratios of the stellar components. The uncertainties on the best-fitted values of $M_*/L_{[3.6]}$ and halo parameters have been derived from the 68\% confidence level.  In most of the plotted RCs the contribution of the dark halo to the velocity curve is above the observed data, so that an unphysically negative $M_*/L_{[3.6]}$ ratio improves the RC fits in almost all galaxies as shown in Figure 1.  This allows us to assess if fits  which are formally good are also physically acceptable.

\subsection{Compliance with SPS models}\label{ml}

In this section we compare the 3.6 micron band $M_*/L$ ratios obtained from RC fits with the independent expectations of stellar population synthesis models. The stellar population synthesis (SPS) models predict a tight linear relation between the color and $M_*/L$ ratio of a stellar population. Redder galaxies should have larger $M_*/L$ ( see, e.g., \citealt{bel01, por04}). The normalization of this relation depends
critically on the shape of the stellar IMF at the low-mass end. These faint stars contribute significantly to the mass, but insignificantly to the luminosity and color of a stellar system (\citealt{bel01}). The slope of this linear relation does not depend on exact details of the history of star formation, i.e. the assumed IMF.

\cite{por04}  suggest the color-$M_*/L_K$ relation using a scaled Salpeter IMF to be

\begin{equation}
 \log(M_*/L_k)= 0.73(B-V)-0.55\label{color}.
\end{equation}

There are other IMF's leading to slightly different relations \citep{Kroupa01, bott97}.  The slope $0.73$ is insensitive to such variations of the IMF, but the $y$-intercept is. The color-$M_*/L_{[3.6]}$ correlation is calculated from Eq. \ref{color} using the following relation \citep{Oh08}

\begin{equation}
 \log(M_*/L_{[3.6]})= 0.92(M_*/L_K)-0.05    \label{3.6-k}.
\end{equation}

\begin{figure}
\includegraphics[width=85mm]{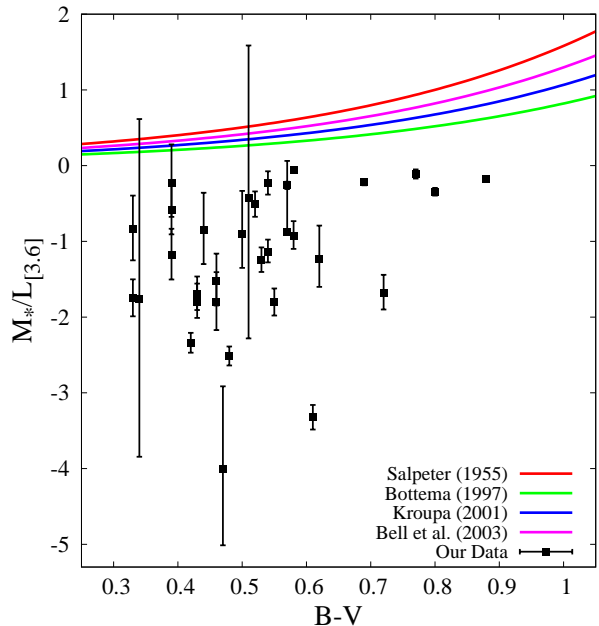}
\caption{A comparison of the best fit stellar $M_*/L_{[3.6]}$ ratios obtained from RC fits with the independent expectations of stellar population synthesis models (lines) versus B-V color. The black filled squares represent the $M_*/L_{[3.6]}$ values for galaxies listed in Table \ref{t1}. Solid lines denote the theoretical predictions of SPS models with different invariant IMFs. These are above the implied $M_*/L_{[3.6]}$ ratios from the RC fits using NFW profiles with $M_{vir}$ and $c$ being free fitting parameters. } \label{f6}
\end{figure}

In Figure \ref{f6} we compare the fitted global disk $M_*/L_{[3.6]}$ ratios to the predictions of SPS models by \cite{bel01} and \cite{por04}. The black symbols are the implied $M_*/L_{[3.6]}$ ratios from the analysis of the RCs. The SPS models of \cite{bel01}, and \cite{por04}, for different IMFs, and for $M_*/L_{[3.6]}$  are also plotted as solid lines.   As can be seen, our results are in significant contradiction to the SPS models. The implied $M_*/L_{[3.6]}$ ratios from the RC fits using NFW profiles as the DM contribution are significantly lower than the SPS models set by different IMFs for stars with masses between 0.1 and 100 $M_{\odot}$.

\begin{figure}
\begin{center}
\includegraphics[width=85mm]{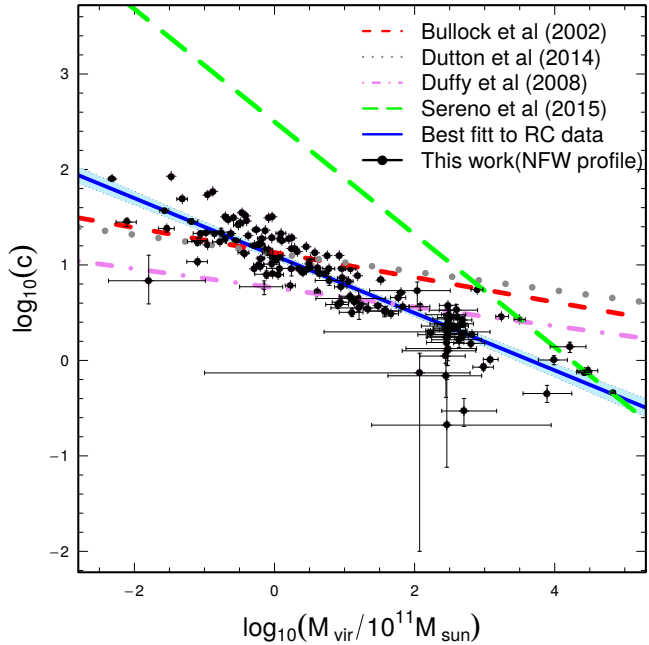}
\caption{$c - M_{vir}$ relation derived from fitting parameters for all galaxies in the SPARC sample. $M_*/L_{[3.6]}$ is a free parameter and is allowed to achieve values less than zero (see Fig \ref{cm-pos} for the cases with positive and negative $M_*/L_{[3.6]}$). The slope of the best fit relation is $\alpha=-0.29 \pm 0.01 $ and the intercept of this relation is $\beta=1.10 \pm 0.02 $. The best-fit relation is represented by the blue line. The $c - M_{vir}$ relation of \citep{bull01} derived from high resolution N-body simulations of a $\Lambda CDM$ cosmological model is represented with the dashed red line. The black dotted line is the $c - M_{vir}$ relation derived by \citep{dutt14} and the violet dotted-dashed line is derived by \citep{duff08} from cosmological simulations. The dashed green line represents the $c-M_{vir}$ relation constrained for dark matter halos of galaxy clusters \citep{ser2015}. See text for more detail.}\label{cm-all}
\end{center}
\end{figure}

\begin{figure}
\begin{center}
\includegraphics[width=85mm]{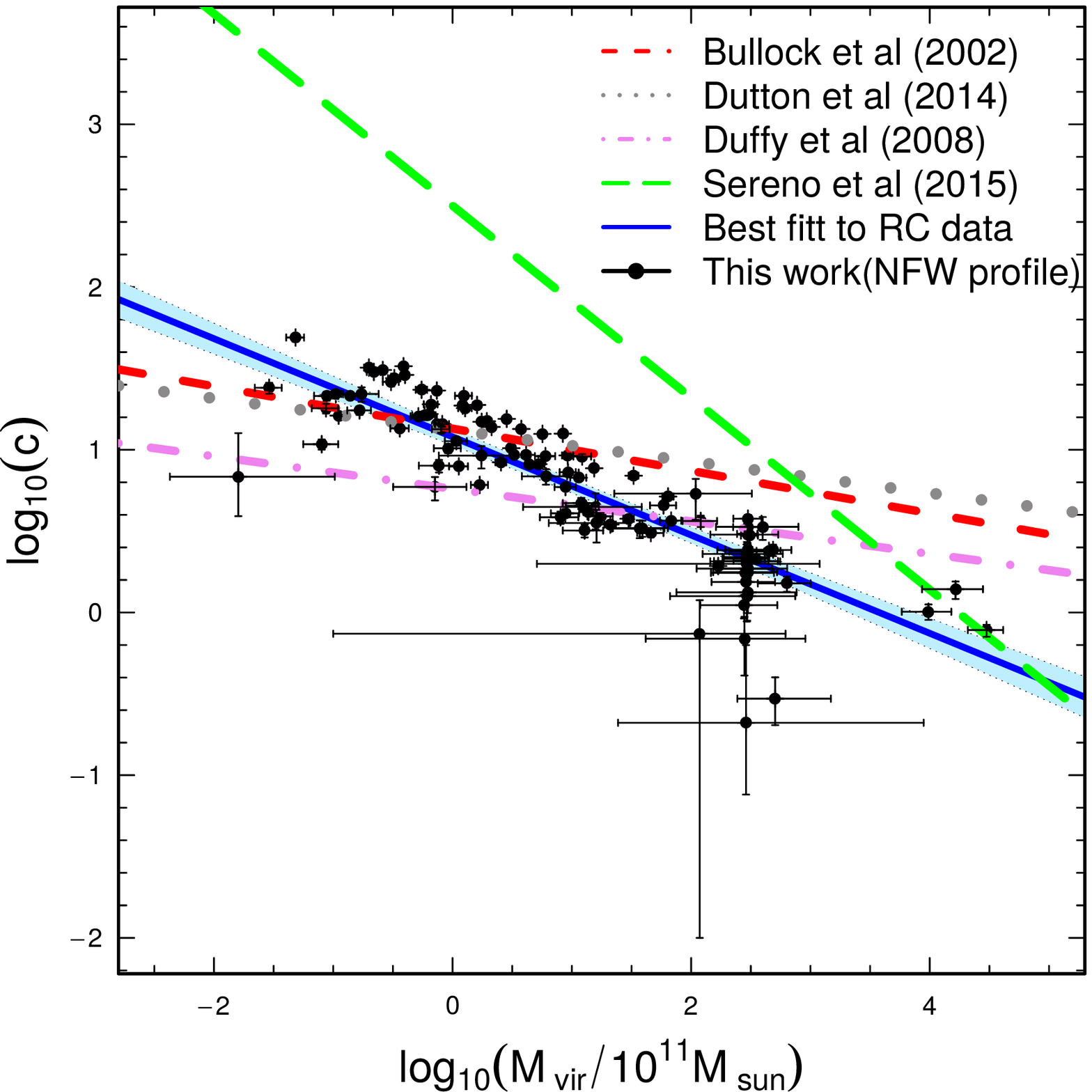}
\includegraphics[width=85mm]{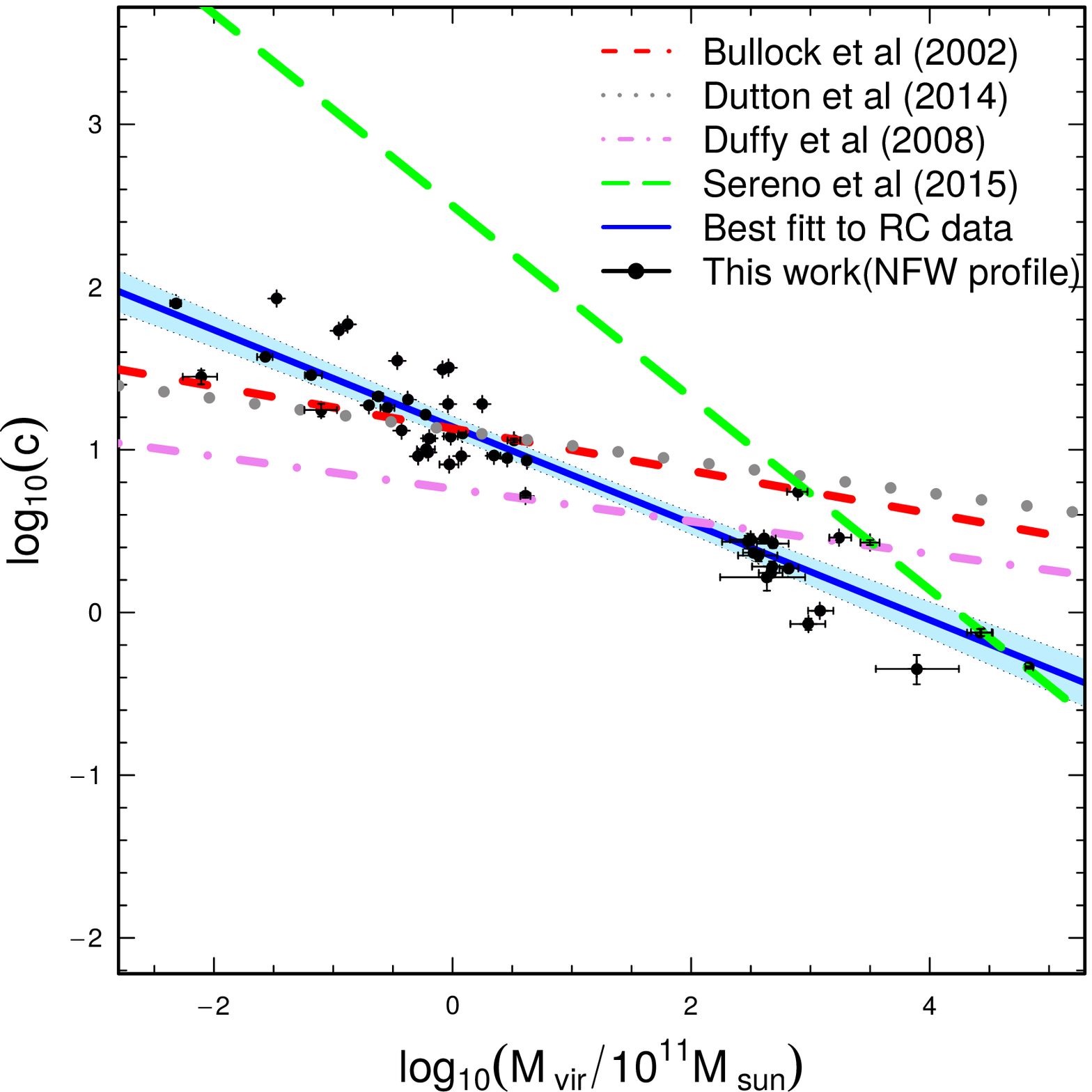}
\caption{ Same as Fig. \ref{cm-all}, but for 120 galaxies that have positive (top) and 51 galaxies with negative (bottom) $M_*/L$ ratios.  The slope of the best fitted $c - M_{vir}$ relation is $\alpha=-0.30 \pm 0.01 $ with the intercept of $\beta=1.08 \pm 0.26 $ for the top panel. The slope of the best fit relation is $\alpha=-0.29 \pm 0.01 $ and an intercept of this relation is $\beta=1.14 \pm 0.27 $ for the bottom panel. }\label{cm-pos}
\end{center}
\end{figure}

\subsection{Comparison with cosmological relations}\label{cm}

The Newtonian N-body technique has allowed us to follow the detailed hierarchical build-up of virialised DM structures, resulting in near spherical haloes that are well described by the NFW profile.  Cosmological simulations of structure formation find that virial masses, $M_{vir}$, and concentrations, $c$, of NFW-like DM haloes are correlated as in Eq.~\ref{eq:c-mvir}, with the average concentration of a halo being a weakly decreasing function of $M_{vir}$.

From the theoretical side we know that  \textbf{i)} halo structure is sensitive to cosmological parameters \citep{macc08} and \textbf{ii)} the galaxy formation process can cause haloes to both contract or expand \citep{dici14}.

\cite{bull01}, \cite{duff08} and \cite{dutt14} have used   a set of N-body simulations to constrain the effects of the cosmological parameters on dark-matter halo structure assuming baryons do not contribute, and to quantify the evolution of the structure of CDM haloes as a population across cosmic time. \cite{bull01} assumed a flat $\Lambda$CDM cosmology and \cite{duff08} applied  a  set  of large  N-body simulations  to  investigate  the  dependence  of  dark matter  halo concentrations on halo mass in the $\Lambda$CDM cosmological model with Wilkinson Microwave Anisotropy Probe (WMAP) parameters. \cite{dutt14}  then applied the Plank cosmological parameters in N-body simulations. On the other hand, \cite{ser2015}   found   a   relation   between   DM   mass and concentration (the $c-M_{vir}$  relation) of galaxy clusters using weak lensing data. Assuming a linear $\log(c)-\log(M_{vir})$  relation as follow,

\begin{equation}
 \log(c)=\alpha\log(M_{vir})+\beta\label{c-m},
\end{equation}

we tabulate the best-fitting coefficient for different analysis in Table \ref{t2}.

In Fig \ref{cm-all} we display the derived $c-M_{vir}$ relationship at redshift z=0 from our RC fits  of all  galaxies in the SPARC sample. The galaxies are shown as black symbols and the error bars denote 68\% confidence intervals.  A clear correlation is seen between these two parameters. Low concentrations correspond to  high virial masses.  Overlaid are $c-M_{vir}$ relations from different cosmological analysis and simulations.  The results of the RC fits strongly disagree with those predicted from different cosmological simulations which are flatter such that many of the galaxies with large $M_{vir}$ are significantly less-concentrated than expected from cosmological $\Lambda$CDM simulations.

We quantify the  goodness-of-fit by computing the standard error, SE, defined as

\begin{equation}
 SE=\frac{1}{(N-1)}\sum_{i=1}^{N}(v_{i}-v_{model})^2\label{SE},
\end{equation}
\noindent where $N$ is the number of galaxies in our sample. The SE values for different cosmological analysis are given in Table \ref{t2}.

As shown in Fig. \ref{cm-pos}, excluding the data of galaxies with the negative $M_*/L$ ratio does not remedy the case and the $c-M_{vir}$ relation that emerges from the RC analysis mismatches those predicted from different cosmological analysis.

\section{Conclusions} \label{conclusion}

Although it is  widely accepted that the $\Lambda$CDM hypothesis is successful in explaining the observed data on cosmological scales (but see also Kroupa 2012; 2015),  it appears to fail rather significantly in reproducing the observed data on galactic and subgalactic scales (Kroupa et al. 2010).  In this paper we assessed if this incompatibility existes in rotation curves of galaxies by calculating dark-matter halo fits for a sample of 171 galaxies from the SPARC dataset. The selected galaxies in the sample cover a large range of luminosities and morphological types. We used the cosmological motivated NFW halo as one of the most commonly used dark matter halo profiles to explain the dynamics of observed disk galaxies. Our least square fit results are:

\begin{itemize}
\item  Although we can explain RC observations of a large sample of galaxies (including 171 galactic RCs) by adding the two-component NFW dark-halo models (three free parameters) into the baryonic mass distribution, we do not recover the cosmological  $c-M_{vir}$ relationship. The RC analysis predicts a significantly steeper $c-M_{vir}$ relationship than we expect from cosmological simulations. According to the RC analysis we observe much lower/higher concentrations than cosmological simulations in high/low mass galaxies, respectively.

\item Moreover, in 51 cases (about 30\% of our sample), the best fitting RCs lead to unphysically negative global 3.6 micron band $M_*/L$ ratios which are clearly at odds with stellar population synthesis (SPS) models. Indeed, an impressive result from the dark-halo fits presented here is that the color-$M_*/L_{[3.6]}$ relation of the stellar populations in the galaxies required to fit the observed rotation curves are inconsistent with those of SPS models, providing a further test that of the $\Lambda$CDM hypothesis for reproducing the dynamics of galaxies fails.
\end{itemize}

It is important to remember that the theoretical prediction of dark matter cusps relies on pure collisionless CDM structure formation simulations (i.e., stars and gas have no impact on the underlying dark matter distribution). The density profiles of DM haloes can be affected by various baryonic processes. Although the supernova feedback could impulsively drive gas out of a galaxy, transforming a central cusp to a core \citep{nav96}, \cite{Gnedin02} found that the effect is very small.  However, \cite{Read05} argued that multiple repeated bursts can cause this small effect to accumulate, gradually grinding a dark matter cusp down to a core.  The most recently used IllustrisTNG (TNG) high resolution hydrodynamic cosmological simulations provide theoretical expectations for the DM mass fractions within the inner regions of haloes \citep{Lovell18}. They found that baryons can pull more DM into the centre of the galaxy as the gas cools and condenses. This enhancement leads TNG present-day galaxies to be dominated by DM within their inner regions meaning that the models are even worse than assumed here in terms of the DM content when baryons are taken into account. In any case, observational indications that blow out of gas even in star-bursting dwarf galaxies does not occur \citep{Lelli14, Concas17}.

It should be noted that our conclusion is based on using the NFW halo profile and it remains to be seen whether other modified dark matter halo profiles, such as the baryonic-mass dependent halo mass profile proposed by \cite{dici14}, can remedy the discrepancy between the $c-M_{vir}$ relation resulting from RC fits with those from cosmological analysis that has been discussed by some authors \citep{katz16, pace16}.

By means of high resolution cosmological hydrodynamic simulations, including the effects of baryonic processes on their host DM haloes, \cite{dici14} proposed a stellar mass dependent density profile for the DM distribution within a galaxy (hereafter, the DC14 profile). They found that the best fit parameters of the DM density profiles vary as a function of the stellar-to-halo mass ratio of each galaxy. Most recently, \cite{katz16} showed that the DC14 halo profiles provide better fits to the data than those of the NFW profile. On the other hand, Wu and Kroupa (2015) have shown that the best models of galaxy formation and evolution compared in the dark matter framework fail to reproduce observed galaxies.

The results found here based on NFW halo profiles indicate a serious problem for the dark matter models. It is interesting to note that more direct tests for the presence of dark matter through the process of dynamical friction, based on Milky Way satellite galaxies \citep{Angus11} and an analysis of the M81 group of galaxies \citep{Oehm17} are also indicating a major tension with the presence of dark matter halos made of particles. More research is useful to further test the notion that the internal dynamics of galaxies is significantly affected by dark matter halos made of particles.

\begin{table*}
	\begin{center}
		\begin{tabular}{|l|c|l|c|c|c|c|l|l|l|r|} \hline
			$Galaxy$ & $Type$ & $L_{[3.6]}$ & $R_{eff}$ & $R_{disk}$ & $M_{[gas]}$ & $B-V$ & $M_*/L_{[3.6]}$ & $M_{vir}$ & $c$ & $\chi^2_{\nu}$ \\
			& & ($10^9L_\odot$)& $(kpc)$& $(kpc)$& ($10^9M_\odot$)& & & ($10^{10}M_\odot$)& &  \\ \hline
			
			CamB & 10 & 0.075$\pm$0.003 & 1.21 & 0.47 & 0.01 & 0.54 &$-1.14_{-0.14}^{+0.16}$ & $11978.2_{-2091.4}^{+3069.9}$ & $1.02_{-0.06}^{+0.06}$ & 1.27 \\
			D564-8 & 10 & 0.033$\pm$0.004 & 0.72 & 0.61 & 0.03 &  & $-0.40_{-0.58}^{+0.64}$ & $9624.4_{-2480.4}^{+3118.0}$ & $0.85_{-0.07}^{+0.07}$ & 0.76 \\
			D631-7 & 10 & 0.196$\pm$0.009 & 1.22 & 0.70 & 0.29 &  & $-1.18_{-0.41}^{+0.46}$ & $4662.4_{-843.7}^{+998.4}$ & $1.75_{-0.09}^{+0.09}$ & 1.65 \\
			DDO154 & 10 & 0.053$\pm$0.002 & 0.65 & 0.37 & 0.28 & 0.33 & $-1.75_{-0.24}^{+0.25}$ & $5.14_{-0.11}^{+0.11}$ & $9.11_{-0.06}^{+0.06}$ & 2.19 \\
			DDO161 & 10 & 0.548$\pm$0.015 & 2.04 & 1.22 & 1.38 & 0.33 & $-0.84_{-0.41}^{+0.44}$ & $40.88_{-3.30}^{+3.49}$ & $5.20_{-0.13}^{+0.13}$ & 0.23 \\
			DDO168 & 10 & 0.191$\pm$0.005 & 1.29 & 1.02 & 0.41 & 0.32 & $-5.88_{-0.40}^{+0.42}$ & $27.41_{-1.85}^{+1.98}$ & $8.59_{-0.14}^{+0.14}$ & 6.71 \\
			DDO170 & 10 & 0.543$\pm$0.030 & 3.03 & 1.95 & 0.74 &  & $-1.60_{-0.30}^{+0.31}$ & $6.25_{-0.26}^{+0.27}$ & $9.63_{-0.14}^{+0.14}$ & 2.38 \\
			ESO444-G084 & 10 & 0.071$\pm$0.003 & 0.75 & 0.46 & 0.14 &  & $-1.00_{-0.91}^{+0.95}$ & $6.37_{-0.61}^{+0.66}$ & $11.75_{-0.30}^{+0.29}$ & 0.69 \\
			F561-1 & 9 & 4.077$\pm$0.327 & 5.39 & 2.79 & 1.62 &  & $-0.74_{-0.18}^{+0.20}$ & $5.99_{-0.84}^{+0.95}$ & $10.07_{-0.47}^{+0.47}$ & 0.29 \\
			F565-V2 & 10 & 0.559$\pm$0.098 & 3.57 & 2.17 & 0.70 & 0.51 & $-0.43_{-1.85}^{+2.02}$ & $3678_{-1053}^{+1349}$ & $2.24_{-0.18}^{+0.18}$ & 0.46 \\
			F571-8 & 5 & 10.164$\pm$0.412 & 1.40 & 3.56 & 1.78 &  & $-0.17_{-0.06}^{+0.07}$ & $32.69_{-3.30}^{+3.59}$ & $11.42_{-0.32}^{+0.32}$ & 0.64 \\
			IC2574 & 9 & 1.016$\pm$0.012 & 3.18 & 2.78 & 1.04 & 0.42 & $-2.34_{-0.13}^{+0.13}$ & $675970_{-38513}^{+53358}$ & $0.46_{-0.01}^{+0.01}$ & 1.45 \\
			IC4202 & 4 & 179.749$\pm$3.311 & 8.55 & 4.78 & 12.33 &   & $-0.93_{-0.04}^{+0.04}$ & $9.25_{-0.28}^{+0.29}$ & $31.92_{-0.29}^{+0.29}$ & 10.64 \\
			NGC0055 & 9 & 4.628$\pm$0.013 & 3.67 & 6.11 & 1.57 & 0.55 & $-1.80_{-0.18}^{+0.18}$ & $28.72_{-2.00}^{+2.14}$ & $8.86_{-0.17}^{+0.17}$ & 0.13 \\
			NGC0891 & 3 & 138.34$\pm$0.255 & 3.68 & 2.55 & 4.46 & 0.88 & $-0.17_{-0.02}^{+0.02}$ & $0.33_{-0.01}^{+0.01}$ & $84.98_{-0.72}^{+0.72}$ & 2.36 \\
			NGC2366 & 10 & 0.236$\pm$0.005 & 1.08 & 0.65 & 0.65 & 0.39 & $-0.58_{-0.33}^{+0.35}$ & $9.37_{-1.39}^{+1.55}$ & $8.12_{-0.33}^{+0.33}$ & 2.64 \\
			NGC2915 & 11 & 0.641$\pm$0.008 & 0.53 & 0.55 & 0.51 & 0.57 & $-0.88_{-\cdots}^{+0.94}$ & $0.65_{-\cdots}^{+0.13}$ & $28.67_{-\cdots}^{+1.82}$ & 0.31 \\
			NGC2955 & 3 & 319.422$\pm$4.413 & 7.22 & 18.76 & 28.95 & 0.69 & $-0.22_{-0.04}^{+0.04}$ & $1.32_{-0.06}^{+0.06}$ & $58.98_{-0.87}^{+0.87}$ & 1.97 \\
			NGC2998 & 5 & 150.902$\pm$2.085 & 7.06 & 6.20 & 23.45 & 0.57 & $-0.26_{-0.05}^{+0.05}$ & $3.44_{-0.09}^{+0.09}$ & $35.22_{-0.36}^{+0.36}$ &0.73 \\
			NGC3109 & 9 & 0.194$\pm$0.002 & 1.64 & 1.56 & 0.48 & 0.47 & $-4.01_{-1.01}^{+1.09}$ & $270085_{-38650}^{+51522}$ & $0.76_{-0.04}^{+0.04}$ &2.28 \\
			NGC3877 & 5 & 72.535$\pm$0.401 & 4.39 & 2.53 & 1.48 & 0.80 & $-0.35_{-0.05}^{+0.05}$ & $17.72_{-1.12}^{+1.19}$ & $19.08_{-0.35}^{+0.35}$ &3.10 \\
			NGC4068 & 10 & 0.236$\pm$0.005 & 1.11 & 0.59 & 0.15 &  & $-0.91_{-0.19}^{+0.20}$ & $264691_{-50465}^{+63023}$ & $0.75_{-0.04}^{+0.04}$ & 0.15 \\
			NGC4214 & 10 & 1.141$\pm$0.008 & 0.70 & 0.51 & 0.49 & 0.46 & $-1.52_{-0.33}^{+0.36}$ & $0.05_{-0.00}^{+0.00}$ & $79.22_{-2.66}^{+2.70}$ & 0.03 \\
			NGC4389 & 4 & 21.328$\pm$0.216 & 4.05 & 2.79 & 0.54 &  & $-0.24_{-0.04}^{+0.04}$ & $31586_{-4625}^{+5424}$ & $2.69_{-0.09}^{+0.10}$ & 0.31 \\
			NGC5985 & 3 & 208.728$\pm$1.538 & 10.71 & 7.01 & 11.59 & 0.77 & $-0.11_{-0.06}^{+0.06}$ & $8.23_{-0.28}^{+0.29}$ & $31.15_{-0.38}^{+0.38}$ & 1.49 \\
			NGC6789 & 11 & 0.100$\pm$0.003 & 0.52 & 0.31 & 0.02 &  & $-5.09_{-0.54}^{+0.58}$& $7840.0_{-1127.2}^{+1402.3}$ & $5.52_{-0.15}^{+0.15}$ & 0.01 \\
			UGC00634 & 9 & 2.989$\pm$0.146 & 4.26 & 2.45 & 3.66 & 0.48 &$-2.51_{-0.13}^{+0.13}$ & $5.92_{-0.09}^{+0.09}$ & $16.49_{-0.09}^{+0.09}$ &2.36 \\
			UGC00891 & 9 & 0.374$\pm$0.017 & 1.76 & 1.43 & 0.43 & 0.61 & $-3.32_{-0.16}^{+0.16}$ & $11.86_{-0.24}^{+0.25}$ & $9.14_{-0.05}^{+0.05}$ & 0.54 \\
			UGC02023 & 10 & 1.308$\pm$0.033 & 2.73 & 1.55 & 0.48 & 0.62 & $-1.23_{-0.37}^{+0.44}$ & $3156_{-916}^{+1294}$ & $2.81_{-0.21}^{+0.22}$ & 0.49 \\
			UGC02455 & 10 & 3.649$\pm$0.034 & 1.49 & 0.99 & 0.80 & 0.58 & $-0.06_{-0.04}^{+0.04}$ & $4775_{-1343}^{+1741}$ & $1.92_{-0.15}^{+0.15}$ & 0.67 \\
			UGC03205 & 2 & 113.642$\pm$1.361 & 5.35 & 3.19 & 9.68 &   & $-0.18_{-0.05}^{+0.05}$ & $1.11_{-0.03}^{+0.03}$ & $53.85_{-0.62}^{+0.62}$ & 2.05 \\
			UGC04278 & 7 & 1.307$\pm$0.026 & 2.46 & 2.21 & 1.12 & 0.44 &$-0.85_{-0.45}^{+0.50}$ & $4861.0_{-1138.9}^{+1431.8}$ & $2.65_{-0.16}^{+0.16}$& 0.75 \\
			UGC04499 & 8 & 1.552$\pm$0.043 & 2.69 & 1.73 & 1.10 & 0.72 &$-1.68_{-0.22}^{+0.24}$ & $1.99_{-0.13}^{+0.14}$ & $18.78_{-0.39}^{+0.39}$ &0.62 \\
			UGC05414 & 10 & 1.123$\pm$0.028 & 2.33 & 1.47 & 0.57 &  &$-0.07_{-0.16}^{+0.17}$ & $3351.3_{-546.0}^{+629.2}$ & $2.33_{-0.10}^{+0.10}$ &0.77 \\
			UGC05716 & 9 & 0.588$\pm$0.042 & 1.84 & 1.14 & 1.09 &  & $-0.63_{-0.14}^{+0.14}$& $3.74_{-0.10}^{+0.10}$ & $13.11_{-0.11}^{+0.11}$ & 1.68 \\
			UGC05721 & 7 & 0.531$\pm$0.011 & 0.60 & 0.38 & 0.56 & 0.39 &$-0.22_{-0.45}^{+0.50}$ & $0.27_{-0.04}^{+0.04}$ & $37.30_{-1.79}^{+1.76}$ &0.85 \\
			UGC05999 & 10 & 3.384$\pm$0.231 & 4.83 & 3.22 & 2.02 &  &$-2.99_{-0.32}^{+0.34}$ & $12.07_{-0.76}^{+0.80}$ & $12.54_{-0.25}^{+0.25}$ &1.99 \\
			UGC06628 & 9 & 3.739$\pm$0.076 & 4.14 & 2.82 & 1.50 &  & $-0.52_{-0.25}^{+0.30}$& $0.79_{-0.19}^{+0.24}$ & $17.56_{-1.58}^{+1.60}$ & 0.19 \\
			UGC06818 & 9 & 1.588$\pm$0.057 & 2.12 & 1.39 & 1.08 & 0.43 &$-1.80_{-0.21}^{+0.24}$ & $9.65_{-1.05}^{+1.20}$ & $12.08_{-0.35}^{+0.37}$ &0.69 \\
			UGC06917 & 9 & 6.832$\pm$0.120 & 4.52 & 2.76 & 2.02 & 0.53 &$-1.25_{-0.16}^{+0.17}$ & $4.20_{-0.26}^{+0.27}$ & $20.27_{-0.39}^{+0.39}$ &0.51 \\
			UGC06923 & 10 & 2.890$\pm$0.077 & 1.66 & 1.44 & 0.81 & 0.52 &$-0.51_{-0.16}^{+0.17}$ & $2.85_{-0.32}^{+0.35}$ & $18.08_{-0.60}^{+0.60}$ &0.74 \\
			UGC07232 & 10 & 0.113$\pm$0.002 & 0.46 & 0.29 & 0.05 & 0.58 &$-0.92_{-0.18}^{+0.19}$ & $17345_{-2546}^{+3916}$ & $2.89_{-0.10}^{+0.10}$ &0.01 \\
			UGC07323 & 8 & 4.109$\pm$0.042 & 3.26 & 2.26 & 0.72 & 0.54 &$-0.23_{-0.15}^{+0.16}$ & $4075.6_{-747.9}^{+881.9}$ & $2.85_{-0.13}^{+0.13}$ &0.44 \\
			UGC07399 & 8 & 1.156$\pm$0.024 & 1.27 & 1.64 & 0.75 & 0.39 &$-1.18_{-0.33}^{+0.34}$ & $1.17_{-0.08}^{+0.08}$ & $27.87_{-0.56}^{+0.56}$ &0.83 \\
			UGC07524 & 9 & 2.436$\pm$0.025 & 3.61 & 3.46 & 1.78 & 0.46 &$-1.80_{-0.37}^{+0.39}$ & $22.23_{-2.32}^{+2.55}$ & $9.18_{-0.28}^{+0.28}$ &0.55 \\
			UGC07577 & 10 & 0.045$\pm$0.002 & 0.77 & 0.90 & 0.04 & 0.50 &$-0.91_{-0.44}^{+0.58}$ & $77751_{-38239}^{+80740}$ & $0.45_{-0.09}^{+0.10}$ &0.01 \\
			UGC07608 & 10 & 0.264$\pm$0.012 & 1.60 & 1.50 & 0.54 & 0.34 &$-1.76_{-2.09}^{+2.37}$ & $2904.9_{-998.8}^{+1380.4}$ & $2.73_{-0.25}^{+0.25}$ &0.45\\
			UGC08837 & 10 & 0.501$\pm$0.015 & 2.25 & 1.72 & 0.32 & 0.43 &$-1.69_{-0.21}^{+0.23}$ & $6591.2_{-936.8}^{+1139.2}$ & $1.87_{-0.07}^{+0.07}$ &3.79 \\
			UGC09037 & 6 & 68.614$\pm$1.769 & 5.69 & 4.28 & 19.08 &  &$-0.42_{-0.06}^{+0.06}$ & $9.15_{-0.61}^{+0.65}$ & $19.08_{-0.42}^{+0.42}$ &0.67 \\
			UGC09992 & 10 & 0.336$\pm$0.017 & 1.62 & 1.04 & 0.32 &  &$-0.12_{-0.44}^{+0.49}$ & $0.08_{-0.02}^{+0.02}$ & $28.09_{-2.85}^{+2.72}$ &0.01 \\
			UGCA444 & 10 & 0.012$\pm$0.001 & 0.41 & 0.83 & 0.07 &  & $-2.99_{-7.78}^{+9.26}$& $4314.0_{-2298.9}^{+3865.8}$ & $1.65_{-0.29}^{+0.27}$ & 0.05 \\
			\hline

\end{tabular}
\end{center}
\caption{The results for RC fits with assuming NFW halo profiles for 51 out of 171 galaxies with the negative best-fitting values of the the $3.6$ micron band stellar mass-to-light ratios. The results for galaxies with positive stellar mass-to-light ratios are given in the next pages. Columns~1 and 2 give the galaxy name and type. Columns~3 gives the luminosity of the galaxy in the 3.6 $\mu m $-band. Columns 4 and 5  are the effective radius at [3.6] in Kpc (the radius at which half of the total light of the galaxy is emitted) and exponential disk radius, respectively. Column~6 and 7 give the total gaseous mass and $B-V$ color of each individual galaxy, respectively. The best-fitting stellar mass-to-light ratio in the $3.6$ micron band for disks is given in column 8.  Best-fitted parameters for the NFW dark-halo profile are given in columns 9, and 10.  The error bars are the 68\% confidence level.  The last column gives the corresponding reduced $\chi_{\nu}^2$.}
\label{t1}
\end{table*}

\begin{table*}
	\begin{center}
		\ContinuedFloat
		\begin{tabular}{|l|c|l|c|c|c|c|l|l|l|r|} \hline

			$Galaxy$ & $Type$ & $L_{[3.6]}$ & $R_{eff}$ & $R_{disk}$ & $M_{[gas]}$ & $B-V$ & $M_*/L_{[3.6]}$ & $M_{vir}$ & $c$ & $\chi^2_{\nu}$ \\
			& & ($10^9L_\odot$)& $(kpc)$& $(kpc)$& ($10^9M_\odot$)& & & ($10^{10}M_\odot$)& &  \\ \hline
			
			D512-2 & 10 & 0.325$\pm$ 0.022 & 2.37 & 1.24 & 0.08 &  & $1.59_{-0.28}^{+0.30}$& $2772_{-1404}^{+2086}$ & $1.11_{-0.2}^{+0.18}$ & 0.28 \\
			DDO064 & 10 & 0.157$\pm$ 0.007 & 1.20 & 0.69 & 0.21 & 0.40 & $0.09_{-1.06}^{+1.23}$ & $2911_{-1485}^{+2435}$ & $2.29_{-0.36}^{+0.34}$ & 0.60 \\
			ESO079-G014 & 4 & 51.733$\pm$ 0.524 & 7.23 & 5.08 & 3.14 &  & $0.75_{-0.07}^{+0.07}$ & $3088_{-412}^{+460}$ & $2.97_{-0.10}^{+0.10}$ & 3.84 \\
			ESO116-G012 & 7 & 4.292$\pm$ 0.071 & 2.75 & 1.51 & 1.08 &  & $0.38_{-0.14}^{+0.15}$ & $52.0_{-5.2}^{+5.6}$ & $8.16_{-0.22}^{+0.22}$ & 2.51 \\
			ESO563-G021 & 4 & 311.177$\pm$ 2.579 & 10.59 & 5.45 & 24.30 &  & $0.87_{-0.07}^{+0.06}$ & $2992_{-638}^{+762}$ & $3.77_{-0.24}^{+0.22}$ & 18.05 \\
			F563-1 & 9 & 1.903$\pm$ 0.170 & 4.61 & 3.52 & 3.20 & 0.64 & $0.93_{-2.02}^{+2.38}$ & $60.7_{-19.8}^{+25.7}$ & $6.87_{-0.78}^{+0.74}$ & 1.14 \\
			F563-V1 & 10 & 1.540$\pm$ 0.165 & 5.01 & 3.79 & 0.61 &  & $0.86_{-0.38}^{+0.44}$ & $0.16_{-\cdots}^{+0.72}$ & $6.84_{-\cdots}^{+5.83}$ & 0.26 \\
			F563-V2 & 10 & 2.986$\pm$ 0.267 & 4.49 & 2.43 & 2.17 & 0.51 & $4.40_{-1.39}^{+1.56}$ & $2933_{-2759}^{+8031}$ & $1.99_{-1.12}^{+0.72}$ & 0.40 \\
			F567-2 & 9 & 2.134$\pm$ 0.305 & 5.43 & 3.08 & 2.45 &  & $1.61_{-0.35}^{+0.39}$ & $1179_{-\cdots}^{+4859}$ & $0.74_{-\cdots}^{+0.45}$ & 0.55 \\
			F568-1 & 5 & 6.252$\pm$ 0.564 & 7.00 & 5.18 & 4.50 &  & $1.90_{-1.04}^{+1.14}$ & $2980_{-1360}^{+1992}$ & $3.01_{-0.41}^{+0.38}$ & 0.92 \\
			F568-3 & 7 & 8.346$\pm$ 0.592 & 7.47 & 4.99 & 3.20 & 0.61 & $0.80_{-0.44}^{+0.48}$ & $2993_{-1370}^{+2156}$ & $2.06_{-0.31}^{+0.28}$ & 3.56 \\
			F568-V1 & 7 & 3.825$\pm$ 0.384 & 4.40 & 2.85 & 2.49 & 0.57 & $3.76_{-1.00}^{+1.09}$ & $158_{-108}^{+188}$ & $4.45_{-1.22}^{+0.95}$ & 0.15 \\
			F571-V1 & 7 & 1.849$\pm$ 0.267 & 4.38 & 2.47 & 1.22 & 0.55 & $0.59_{-0.75}^{+0.82}$ & $124_{-31}^{+38}$ & $4.53_{-0.35}^{+0.34}$ & 0.40 \\
			F574-1 & 7 & 6.537$\pm$ 0.596 & 5.87 & 4.46 & 3.52 &  & $1.45_{-0.32}^{+0.34}$ & $120_{-32}^{+39}$ & $4.70_{-0.39}^{+0.36}$ & 1.64 \\
			F574-2 & 9 & 2.877$\pm$ 0.384 & 6.48 & 3.76 & 1.70 &  & $0.03_{-0.29}^{+0.35}$ & $2800_{-2204}^{+5055}$ & $0.69_{-0.29}^{+0.25}$ & 0.12 \\
			F579-V1 & 5 & 11.848$\pm$ 0.742 & 5.76 & 3.37 & 2.25 &  & $0.72_{-0.49}^{+0.53}$ & $1.73_{-0.49}^{+0.57}$ & $22.03_{-2.32}^{+2.15}$ & 0.20 \\
			F583-1 & 9 & 0.986$\pm$ 0.093 & 3.74 & 2.36 & 2.13 & 0.39 & $1.42_{-1.04}^{+1.15}$ & $390_{-149}^{+208}$ & $3.30_{-0.38}^{+0.36}$ & 1.64 \\
			F583-4 & 5 & 1.715$\pm$ 0.185 & 3.31 & 1.93 & 0.64 &  & $0.97_{-0.36}^{+0.39}$ & $2855_{-1232}^{+1743}$ & $1.75_{-0.24}^{+0.22}$ & 0.13 \\
			KK98-251 & 10 & 0.085$\pm$ 0.007 & 1.28 & 1.34 & 0.12 &  & $3.95_{-0.89}^{+0.97}$ & $2884_{-\cdots}^{+85371}$ & $0.21_{-\cdots}^{+0.42}$ & 1.68 \\
			NGC0024 & 5 & 3.889$\pm$ 0.036 & 2.01 & 1.34 & 0.68 & 0.58 & $1.78_{-0.37}^{+0.40}$ & $2971_{-1161}^{+1588}$ & $2.41_{-0.28}^{+0.26}$ & 0.35 \\
			NGC0100 & 6 & 3.232$\pm$ 0.063 & 2.81 & 1.66 & 1.99 & 0.65 & $0.44_{-0.23}^{+0.25}$ & $2971_{-910}^{+1162}$ & $2.43_{-0.21}^{+0.20}$ & 0.84 \\
			NGC0247 & 7 & 7.332$\pm$ 0.027 & 5.87 & 3.74 & 1.75 & 0.56 & $1.30_{-0.13}^{+0.14}$ & $174_{-34}^{+39}$ & $3.90_{-0.21}^{+0.2}$ & 1.65 \\
			NGC0289 & 4 & 72.065$\pm$ 0.465 & 2.69 & 6.74 & 27.47 & 0.73 & $0.63_{-0.13}^{+0.13}$ & $114_{-15}^{+16}$ & $6.73_{-0.32}^{+0.31}$ & 1.92 \\
			NGC0300 & 7 & 2.922$\pm$ 0.008 & 1.77 & 1.75 & 0.94 & 0.59 & $0.59_{-0.26}^{+0.27}$ & $88.40_{-14.57}^{+16.55}$ & $5.90_{-0.28}^{+0.27}$ & 0.68 \\
			NGC0801 & 5 & 312.57$\pm$ 3.455 & 7.76 & 8.72 & 23.20 & 0.87 & $0.62_{-0.04}^{+0.03}$ & $3496_{-530}^{+556}$ & $2.13_{-0.11}^{+0.1}$ & 5.54 \\
			NGC1003 & 6 & 6.820$\pm$ 0.075 & 2.76 & 1.61 & 5.88 & 0.55 & $0.70_{-0.15}^{+0.16}$ & $299_{-26}^{+28}$ & $3.77_{-0.10}^{+0.10}$ & 2.31 \\
			NGC1090 & 4 & 72.045$\pm$ 0.796 & 6.36 & 3.53 & 8.78 & 0.68 & $0.45_{-0.05}^{+0.05}$ & $41.39_{-2.8}^{+2.93}$ & $9.30_{-0.22}^{+0.22}$ & 2.50 \\
			NGC1705 & 11 & 0.533$\pm$ 0.010 & 0.49 & 0.39 & 0.14 & 0.38 & $1.28_{-0.49}^{+0.53}$ & $0.88_{-0.17}^{+0.19}$ & $21.40_{-1.38}^{+1.32}$ & 0.07 \\
			NGC2403 & 6 & 10.041$\pm$ 0.028 & 2.16 & 1.39 & 3.20 & 0.47 & $0.29_{-0.03}^{+0.03}$ & $6.40_{-0.14}^{+0.14}$ & $16.58_{-0.11}^{+0.11}$ & 9.14 \\
			NGC2683 & 3 & 80.415$\pm$ 0.222 & 3.34 & 2.18 & 1.41 & 0.89 & $0.39_{-0.07}^{+0.07}$ & $2.188_{-0.244}^{+0.256}$ & $30.07_{-1.27}^{+1.23}$ & 0.88 \\
			NGC2841 & 3 & 188.121$\pm$ 0.520 & 5.51 & 3.64 & 9.78 & 0.87 & $1.01_{-0.08}^{+0.07}$ & $329.3_{-26}^{+27.3}$ & $6.93_{-0.18}^{+0.18}$ & 1.42 \\
			NGC2903 & 4 & 81.863$\pm$ 0.151 & 4.54 & 2.33 & 2.55 & 0.67 & $0.22_{-0.02}^{+0.02}$ & $2.60_{-0.08}^{+0.08}$ & $30.91_{-0.32}^{+0.32}$ & 5.96 \\
			NGC3198 & 5 & 38.279$\pm$ 0.212 & 5.84 & 3.14 & 10.87 & 0.54 & $0.46_{-0.04}^{+0.04}$ & $31.06_{-1.66}^{+1.72}$ & $10.22_{-0.19}^{+0.18}$ & 1.39 \\
			NGC3521 & 4 & 84.836$\pm$ 0.156 & 2.45 & 2.40 & 4.15 & 0.81 & $0.54_{-0.06}^{+0.06}$ & $3979_{-2111}^{+3274}$ & $3.36_{-0.57}^{+0.5}$ & 0.23 \\
			NGC3726 & 5 & 70.234$\pm$ 0.388 & 7.52 & 3.40 & 6.47 & 0.49 & $0.51_{-0.05}^{+0.06}$ & $2977_{-772}^{+951}$ & $2.37_{-0.19}^{+0.18}$ & 2.25 \\
			NGC3741 & 10 & 0.028$\pm$ 0.001 & 0.32 & 0.20 & 0.18 & 0.30 & $0.44_{-0.90}^{+1.15}$ & $88.31_{-18.8}^{+22.65}$ & $4.07_{-0.24}^{+0.23}$ & 0.36 \\
			NGC3769 & 3 & 18.679$\pm$ 0.189 & 2.18 & 3.38 & 5.53 & 0.64 & $0.15_{-0.11}^{+0.12}$ & $6.15_{-0.78}^{+0.85}$ & $16.25_{-0.73}^{+0.72}$ & 0.47 \\
			NGC3893 & 5 & 58.525$\pm$ 0.377 & 2.43 & 2.38 & 5.80 & 0.56 & $0.20_{-0.06}^{+0.07}$ & $3.25_{-0.34}^{+0.36}$ & $27.55_{-0.97}^{+0.95}$ & 0.53 \\
			NGC3917 & 6 & 21.966$\pm$ 0.202 & 5.56 & 2.63 & 1.89 & 0.72 & $1.04_{-0.10}^{+0.10}$ & $2980_{-963}^{+1227}$ & $2.14_{-0.21}^{+0.19}$ & 2.89 \\
			NGC3949 & 4 & 38.067$\pm$ 0.280 & 2.39 & 3.59 & 3.37 & 0.45 & $0.13_{-0.05}^{+0.05}$ & $3.05_{-0.39}^{+0.42}$ & $26.13_{-0.98}^{+0.95}$ & 0.14 \\
			NGC3953 & 4 & 141.301$\pm$ 0.521 & 6.17 & 4.89 & 2.83 & 0.77 & $0.35_{-0.04}^{+0.04}$ & $5.50_{-0.52}^{+0.54}$ & $23.37_{-0.74}^{+0.72}$ & 0.27 \\
			NGC3992 & 4 & 226.932$\pm$ 0.836 & 9.99 & 4.96 & 16.60 & 0.77 & $0.47_{-0.06}^{+0.06}$ & $16.04_{-1.01}^{+1.04}$ & $18.86_{-0.45}^{+0.44}$ & 0.48 \\
			NGC4010 & 7 & 17.193$\pm$ 0.190 & 6.63 & 2.81 & 2.83 & 0.54 & $0.27_{-0.11}^{+0.12}$ & $586_{-117}^{+137}$ & $4.55_{-0.24}^{+0.23}$ & 2.32 \\
			NGC4013 & 3 & 79.094$\pm$ 0.364 & 4.11 & 3.53 & 2.97 & 0.96 & $0.60_{-0.05}^{+0.06}$ & $4419_{-760}^{+1023}$ & $2.40_{-0.14}^{+0.11}$ & 0.78 \\
			NGC4051 & 4 & 95.268$\pm$ 0.439 & 6.87 & 4.65 & 2.70 & 0.65 & $0.36_{-0.05}^{+0.06}$ & $5.83_{-1.08}^{+1.19}$ & $16.29_{-1.03}^{+0.97}$ & 0.80 \\
			NGC4085 & 5 & 21.724$\pm$ 0.200 & 2.00 & 1.65 & 1.35 & 0.58 & $0.06_{-0.05}^{+0.05}$ & $122_{-17}^{+20}$ & $9.09_{-0.31}^{+0.31}$ & 4.01 \\
			NGC4088 & 4 & 107.286$\pm$ 0.494 & 6.10 & 2.58 & 8.23 & 0.59 & $0.38_{-0.05}^{+0.05}$ & $97590_{-34507}^{+45429}$ & $1.01_{-0.12}^{+0.11}$ & 0.45 \\
			NGC4100 & 4 & 59.394$\pm$ 0.328 & 4.93 & 2.15 & 3.10 & 0.73 & $0.51_{-0.07}^{+0.07}$ & $6.59_{-0.74}^{+0.78}$ & $19.03_{-0.76}^{+0.73}$ & 0.90 \\
			NGC4138 & 0 & 44.111$\pm$ 0.284 & 1.91 & 1.51 & 1.48 & 0.84 & $0.35_{-0.10}^{+0.10}$ & $0.483_{-0.07}^{+0.074}$ & $48.97_{-2.78}^{+2.65}$ & 0.96 \\
			NGC4157 & 3 & 105.620$\pm$ 0.486 & 4.44 & 2.32 & 8.23 & 0.80 & $0.49_{-0.07}^{+0.06}$ & $4861_{-1394}^{+1733}$ & $2.42_{-0.21}^{+0.20}$ & 0.34 \\
			NGC4183 & 6 & 10.838$\pm$ 0.150 & 4.47 & 2.79 & 3.51 & 0.55 & $0.60_{-0.24}^{+0.26}$ & $10.76_{-1.87}^{+2.06}$ & $11.30_{-0.70}^{+0.67}$ & 0.17 \\
			NGC4217 & 3 & 85.299$\pm$ 0.393 & 5.28 & 2.94 & 2.56 & 0.87 & $0.10_{-0.04}^{+0.04}$ & $12.74_{-1.37}^{+1.48}$ & $18.03_{-0.62}^{+0.51}$ & 3.50 \\
			NGC4559 & 6 & 19.377$\pm$ 0.107 & 3.82 & 2.10 & 5.81 & 0.45 & $0.12_{-0.11}^{+0.12}$ & $8.20_{-0.91}^{+0.97}$ & $14.38_{-0.56}^{+0.55}$ & 0.20 \\
			NGC5005 & 4 & 178.720$\pm$ 0.494 & 3.97 & 9.45 & 1.28 & 0.80 & $0.50_{-0.08}^{+0.09}$ & $1092_{-859}^{+1742}$ & $5.38_{-1.67}^{+1.24}$ & 0.06 \\
			NGC5033 & 5 & 110.509$\pm$ 0.407 & 2.94 & 5.16 & 11.31 & 0.55 & $0.30_{-0.04}^{+0.04}$ & $7.383_{-0.249}^{+0.253}$ & $23.12_{-0.29}^{+0.29}$ & 2.67 \\
			NGC5055 & 4 & 152.922$\pm$ 0.282 & 4.18 & 3.20 & 11.72 & 0.72 & $0.30_{-0.01}^{+0.01}$ & $21.23_{-0.43}^{+0.43}$ & $13.81_{-0.10}^{+0.09}$ & 2.56 \\
			NGC5371 & 4 & 340.393$\pm$ 1.881 & 9.80 & 7.44 & 11.18 & 0.70 & $0.22_{-0.02}^{+0.02}$ & $4.00_{-0.17}^{+0.18}$ & $28.84_{-0.47}^{+0.47}$ & 4.25 \\
			NGC5585 & 7 & 2.943$\pm$ 0.033 & 2.27 & 1.53 & 1.68 & 0.46 & $0.13_{-0.07}^{+0.07}$ & $25.48_{-2.26}^{+2.38}$ & $8.38_{-0.21}^{+0.20}$ & 5.83 \\  \hline
			
		\end{tabular}
		\caption{Continued}
	\end{center}
	
\end{table*}

\begin{table*}
	\begin{center}
		\ContinuedFloat
		\begin{tabular}{|l|c|l|c|c|c|c|l|l|l|r|} \hline
			
			$Galaxy$ & $Type$ & $L_{[3.6]}$ & $R_{eff}$ & $R_{disk}$ & $M_{[gas]}$ & $B-V$ & $M_*/L_{[3.6]}$ & $M_{vir}$ & $c$ & $\chi^2_{\nu}$ \\
			& & ($10^9L_\odot$)& $(kpc)$& $(kpc)$& ($10^9M_\odot$)& & & ($10^{10}M_\odot$)& &  \\ \hline
			
			NGC5907 & 5 & 175.425$\pm$ 0.646 & 7.88 & 5.34 & 21.03 & 0.78 & $0.05_{-0.03}^{+0.03}$ & $3.88_{-0.09}^{+0.09}$ & $32.61_{-0.30}^{+0.30}$ & 3.97 \\
			NGC6015 & 6 & 32.129$\pm$ 0.237 & 3.92 & 2.30 & 5.83 & 0.57 & $0.92_{-0.04}^{+0.05}$ & $679_{-114}^{+125}$ & $3.67_{-0.18}^{+0.17}$ & 8.14 \\
			NGC6195 & 3 & 391.076$\pm$ 6.123 & 9.52 & 13.94 & 20.91 & 0.75 & $0.62_{-0.03}^{+0.03}$ & $87065128_{-25536972}^{+42879584}$ & $0.09_{-0.02}^{+0.02}$ & 1.43 \\
			NGC6503 & 6 & 12.845$\pm$ 0.059 & 1.62 & 2.16 & 1.74 & 0.68 & $0.39_{-0.05}^{+0.05}$ & $7.39_{-0.25}^{+0.26}$ & $14.42_{-0.17}^{+0.17}$ & 1.44 \\
			NGC6674 & 3 & 214.654$\pm$ 1.977 & 7.75 & 6.04 & 32.17 & 0.84 & $1.23_{-0.04}^{+0.04}$ & $4113745_{-354726}^{+361846}$ & $0.25_{-0.01}^{+0.01}$ & 1.31 \\
			NGC6946 & 6 & 66.173$\pm$ 0.122 & 4.20 & 2.44 & 5.67 & 0.80 & $0.50_{-0.03}^{+0.03}$ & $92.76_{-11.89}^{+12.83}$ & $7.28_{-0.26}^{+0.25}$ & 1.62 \\
			NGC7331 & 3 & 250.631$\pm$ 0.693 & 3.99 & 5.02 & 11.07 & 0.87 & $0.39_{-0.02}^{+0.02}$ & $640.1_{-57.9}^{+61.5}$ & $5.16_{-0.13}^{+0.14}$ & 0.83 \\
			NGC7793 & 7 & 7.050$\pm$ 0.026 & 2.19 & 1.21 & 0.86 & 0.54 & $0.60_{-0.16}^{+0.17}$ & $17.44_{-8.66}^{+12.43}$ & $9.21_{-1.55}^{+1.32}$ & 0.90 \\
			NGC7814 & 2 & 74.529$\pm$ 0.343 & 2.08 & 2.54 & 1.07 & 0.99 & $0.52_{-0.05}^{+0.05}$ & $12.158_{-0.743}^{+0.768}$ & $18.70_{-0.38}^{+0.37}$ & 0.45 \\
			PGC51017 & 11 & 0.155$\pm$ 0.014 & 1.28 & 0.53 & 0.20 &  & $0.06_{-0.16}^{+0.17}$ & $0.00040_{-0.00009}^{+0.0001}$ & $125.08_{-13.40}^{+12.15}$ & 0.01 \\
			UGC00128 & 8 & 12.020$\pm$ 0.565 & 9.63 & 5.95 & 7.43 & 0.60 & $0.19_{-0.09}^{+0.09}$ & $32.82_{-0.62}^{+0.62}$ & $9.27_{-0.06}^{+0.06}$ & 3.19 \\
			UGC00191 & 9 & 2.004$\pm$ 0.063 & 2.50 & 1.58 & 1.34 & 0.44 & $1.35_{-0.11}^{+0.12}$ & $138_{-12}^{+12}$ & $4.14_{-0.10}^{+0.10}$ & 3.19 \\
			UGC00731 & 10 & 0.323$\pm$ 0.019 & 1.40 & 2.30 & 1.81 &  & $10.22_{-1.61}^{+1.68}$ & $128_{-38}^{+47}$ & $3.19_{-0.30}^{+0.28}$ & 0.08 \\
			UGC01230 & 9 & 7.620$\pm$ 0.379 & 6.45 & 4.34 & 6.43 & 0.54 & $2.44_{-0.85}^{+0.97}$ & $161_{-86}^{+124}$ & $3.55_{-0.86}^{+0.72}$ & 0.91 \\
			UGC01281 & 8 & 0.353$\pm$ 0.009 & 2.01 & 1.63 & 0.29 &  & $0.05_{-1.05}^{+1.22}$ & $2995_{-1272}^{+1935}$ & $2.07_{-0.27}^{+0.26}$ & 1.34 \\
			UGC02259 & 8 & 1.725$\pm$ 0.038 & 2.40 & 1.62 & 0.49 &  & $0.26_{-0.16}^{+0.17}$ & $1.39_{-0.07}^{+0.07}$ & $21.37_{-0.34}^{+0.34}$ & 0.82 \\
			UGC02487 & 0 & 489.955$\pm$ 4.061 & 9.63 & 7.89 & 17.96 &  & $0.90_{-0.04}^{+0.04}$ & $84.011_{-2.596}^{+2.632}$ & $12.56_{-0.15}^{+0.14}$ & 4.32 \\
			UGC02885 & 5 & 403.525$\pm$ 4.088 & 12.20 & 11.4 & 40.08 & 0.47 & $0.88_{-0.09}^{+0.10}$ & $299343_{-79325}^{+95311}$ & $0.78_{-0.07}^{+0.07}$ & 1.43 \\
			UGC02916 & 2 & 124.153$\pm$ 1.830 & 2.80 & 6.15 & 23.27 &  & $0.43_{-0.02}^{+0.02}$ & $19.371_{-1.47}^{+1.526}$ & $14.97_{-0.33}^{+0.32}$ & 10.45 \\
			UGC02953 & 2 & 259.518$\pm$ 0.717 & 5.03 & 3.55 & 7.68 &  & $0.60_{-0.02}^{+0.02}$ & $56.663_{-1.943}^{+1.976}$ & $12.52_{-0.14}^{+0.14}$ & 5.67 \\
			UGC03546 & 1 & 101.336$\pm$ 0.747 & 2.58 & 3.79 & 2.68 & 0.89 & $0.42_{-0.02}^{+0.02}$ & $17.525_{-0.617}^{+0.628}$ & $14.85_{-0.17}^{+0.17}$ & 0.84 \\
			UGC03580 & 1 & 13.266$\pm$ 0.195 & 1.84 & 2.43 & 4.37 &  & $0.23_{-0.04}^{+0.04}$ & $44.120_{-3.029}^{+3.178}$ & $8.11_{-0.17}^{+0.17}$ & 3.12 \\
			UGC04325 & 9 & 2.026$\pm$ 0.035 & 2.79 & 1.86 & 0.68 & 0.44 & $2.24_{-0.22}^{+0.22}$ & $0.87_{-0.18}^{+0.20}$ & $17.98_{-1.28}^{+1.18}$ & 1.96 \\
			UGC04483 & 10 & 0.013$\pm$ 0.001 & 0.26 & 0.18 & 0.03 &  & $0.29_{-0.59}^{+0.66}$ & $0.80_{-0.21}^{+0.26}$ & $10.82_{-0.80}^{+0.77}$ & 0.71 \\
			UGC05005 & 10 & 4.100$\pm$ 0.283 & 5.02 & 3.20 & 3.09 &  & $0.07_{-0.65}^{+0.79}$ & $371_{-137}^{+188}$ & $3.28_{-0.42}^{+0.40}$ & 0.22 \\
			UGC05253 & 2 & 171.582$\pm$ 0.790 & 4.28 & 8.07 & 16.40 & 0.74 & $0.63_{-0.02}^{+0.02}$ & $92.25_{-8.25}^{+8.69}$ & $9.23_{-0.25}^{+0.25}$ & 4.15 \\
			UGC05764 & 10 & 0.085$\pm$ 0.006 & 1.20 & 1.17 & 0.16 & 0.53 & $3.26_{-0.46}^{+0.46}$ & $1.11_{-0.05}^{+0.05}$ & $16.21_{-0.20}^{+0.20}$ & 7.53 \\
			UGC05829 & 10 & 0.564$\pm$ 0.019 & 2.91 & 1.99 & 1.02 & 0.21 & $0.76_{-1.07}^{+1.18}$ & $2997_{-1302}^{+1874}$ & $1.77_{-0.24}^{+0.23}$ & 0.09 \\
			UGC05918 & 10 & 0.233$\pm$ 0.011 & 2.63 & 1.66 & 0.30 & 0.54 & $2.74_{-0.99}^{+1.08}$ & $7.12_{-3.49}^{+4.98}$ & $5.91_{-1.03}^{+0.89}$ & 0.10 \\
			UGC05986 & 9 & 4.695$\pm$ 0.048 & 3.12 & 1.67 & 2.67 & 0.42 & $0.93_{-0.12}^{+0.13}$ & $2971_{-474}^{+530}$ & $3.04_{-0.12}^{+0.12}$ & 5.52 \\
			UGC06399 & 9 & 2.296$\pm$ 0.072 & 3.45 & 2.05 & 0.67 &  & $1.25_{-0.33}^{+0.35}$ & $2971_{-1001}^{+1299}$ & $2.19_{-0.21}^{+0.2}$ & 0.57 \\
			UGC06446 & 7 & 0.988$\pm$ 0.032 & 2.06 & 1.49 & 1.38 & 0.39 & $1.39_{-0.65}^{+0.69}$ & $9.16_{-1.90}^{+2.19}$ & $10.21_{-0.67}^{+0.64}$ & 0.21 \\
			UGC06614 & 1 & 124.350$\pm$ 2.520 & 3.68 & 5.10 & 21.89 & 0.87 & $0.46_{-0.10}^{+0.10}$ & $1203_{-311}^{+378}$ & $3.65_{-0.31}^{+0.29}$ & 0.28 \\
			UGC06667 & 6 & 1.397$\pm$ 0.066 & 3.50 & 5.15 & 0.81 & 0.65 & $8.82_{-1.14}^{+1.18}$ & $7.70_{-2.16}^{+2.53}$ & $7.99_{-0.74}^{+0.68}$ & 1.16 \\
			UGC06786 & 0 & 73.407$\pm$ 0.676 & 3.42 & 3.60 & 5.03 & 0.85 & $0.60_{-0.04}^{+0.04}$ & $37.709_{-1.752}^{+1.805}$ & $13.42_{-0.20}^{+0.20}$ & 1.21 \\
			UGC06787 & 2 & 98.256$\pm$ 0.543 & 2.88 & 5.37 & 5.03 & 0.90 & $0.54_{-0.03}^{+0.03}$ & $28.45_{-1.12}^{+1.15}$ & $15.49_{-0.19}^{+0.19}$ & 25.90 \\
			UGC06930 & 7 & 8.932$\pm$ 0.140 & 4.38 & 3.94 & 3.23 & 0.59 & $0.10_{-0.26}^{+0.27}$ & $8.01_{-1.13}^{+1.23}$ & $13.38_{-0.65}^{+0.63}$ & 0.29 \\
			UGC06973 & 2 & 53.870$\pm$ 0.347 & 1.61 & 1.07 & 1.75 & 1.00 & $0.10_{-0.02}^{+0.02}$ & $1.989_{-0.155}^{+0.162}$ & $31.99_{-0.79}^{+0.77}$ & 0.54 \\
			UGC06983 & 6 & 5.298$\pm$ 0.102 & 3.95 & 3.21 & 2.97 & 0.45 & $0.13_{-0.29}^{+0.31}$ & $5.17_{-0.63}^{+0.70}$ & $16.05_{-0.66}^{+0.66}$ & 0.62 \\
			UGC07089 & 8 & 3.585$\pm$ 0.089 & 3.90 & 2.26 & 1.21 &  & $0.19_{-0.23}^{+0.25}$ & $2951_{-952}^{+1255}$ & $2.06_{-0.19}^{+0.19}$ & 0.22 \\
			UGC07125 & 9 & 2.712$\pm$ 0.080 & 3.92 & 3.38 & 4.63 &  & $0.29_{-0.19}^{+0.19}$ & $16.88_{-2.3}^{+2.49}$ & $6.08_{-0.29}^{+0.28}$ & 0.91 \\
			UGC07151 & 6 & 2.284$\pm$ 0.025 & 2.17 & 1.25 & 0.62 & 0.43 & $1.15_{-0.13}^{+0.13}$ & $6343_{-2394}^{+3019}$ & $1.51_{-0.17}^{+0.16}$ & 2.06 \\
			UGC07261 & 8 & 1.753$\pm$ 0.048 & 2.66 & 1.20 & 1.39 &  & $0.30_{-0.28}^{+0.30}$ & $1.66_{-0.31}^{+0.35}$ & $17.46_{-1.09}^{+1.05}$ & 0.02 \\
			UGC07559 & 10 & 0.109$\pm$ 0.004 & 0.98 & 0.58 & 0.17 & 0.34 & $0.01_{-0.44}^{+0.5}$ & $2877_{-1235}^{+1791}$ & $1.54_{-0.21}^{+0.20}$ & 0.40 \\
			UGC07603 & 7 & 0.376$\pm$ 0.008 & 0.85 & 0.53 & 0.26 &  & $0.07_{-0.32}^{+0.34}$ & $3.61_{-0.52}^{+0.59}$ & $13.54_{-0.55}^{+0.55}$ & 1.58 \\
			UGC07690 & 10 & 0.858$\pm$ 0.018 & 0.86 & 0.57 & 0.39 & 0.47 & $0.61_{-0.20}^{+0.21}$ & $0.29_{-0.06}^{+0.07}$ & $24.04_{-1.92}^{+1.79}$ & 0.41 \\
			UGC07866 & 10 & 0.124$\pm$ 0.004 & 0.95 & 0.61 & 0.12 & 0.26 & $0.85_{-0.59}^{+0.68}$ & $2988_{-2031}^{+3811}$ & $1.33_{-0.35}^{+0.31}$ & 0.04 \\
			UGC08286 & 6 & 1.255$\pm$ 0.018 & 2.25 & 1.05 & 0.64 & 0.53 & $2.37_{-0.24}^{+0.25}$ & $2926_{-574}^{+673}$ & $2.08_{-0.11}^{+0.11}$ & 1.73 \\
			UGC08490 & 9 & 1.017$\pm$ 0.012 & 1.14 & 0.67 & 0.72 & 0.41 & $0.47_{-0.40}^{+0.43}$ & $1.05_{-0.13}^{+0.14}$ & $21.95_{-0.97}^{+0.94}$ & 0.12 \\
			UGC08550 & 7 & 0.289$\pm$ 0.009 & 1.01 & 0.45 & 0.29 & 1.26 & $1.17_{-0.33}^{+0.34}$ & $11.28_{-1.68}^{+1.86}$ & $7.91_{-0.34}^{+0.33}$ & 0.58 \\
			UGC08699 & 2 & 50.302$\pm$ 0.695 & 1.91 & 3.09 & 3.74 & 0.97 & $0.70_{-0.05}^{+0.05}$ & $60.02_{-9.56}^{+10.6}$ & $9.14_{-0.44}^{+0.42}$ & 0.76 \\
			UGC09133 & 2 & 282.926$\pm$ 2.345 & 5.92 & 6.97 & 33.43 & 0.87 & $0.61_{-0.03}^{+0.03}$ & $154.00_{-5.48}^{+5.57}$ & $7.70_{-0.10}^{+0.10}$ & 7.55 \\
			UGC10310 & 9 & 1.741$\pm$ 0.053 & 3.12 & 1.80 & 1.20 & 0.42 & $2.04_{-0.37}^{+0.39}$ & $2944_{-2068}^{+3720}$ & $1.26_{-0.37}^{+0.30}$ & 0.15 \\
			UGC11455 & 6 & 374.322$\pm$ 3.792 & 10.06 & 5.93 & 13.34 &  & $0.51_{-0.05}^{+0.05}$ & $2994_{-648}^{+768}$ & $3.35_{-0.22}^{+0.21}$ & 5.16 \\
			UGC11557 & 8 & 12.101$\pm$ 0.212 & 4.18 & 2.75 & 2.61 &  & $0.23_{-0.13}^{+0.15}$ & $2971_{-1662}^{+2810}$ & $1.86_{-0.36}^{+0.33}$ & 1.27 \\
			UGC11820 & 9 & 0.970$\pm$ 0.047 & 2.74 & 2.08 & 1.98 &  & $1.64_{-0.27}^{+0.27}$ & $1684_{-117}^{+124}$ & $1.93_{-0.04}^{+0.04}$ & 0.40 \\
			UGC11914 & 2 & 150.028$\pm$ 0.553 & 3.12 & 2.44 & 0.89 & 0.90 & $0.82_{-0.03}^{+0.03}$ & $164651_{-68802}^{+96866}$ & $1.39_{-0.18}^{+0.17}$ & 5.11 \\  \hline
			
		\end{tabular}
		\caption{Continued}
	\end{center}
	
\end{table*}

\begin{table*}
	\begin{center}
		\ContinuedFloat
		\begin{tabular}{|l|c|l|c|c|c|c|l|l|l|r|} \hline
			
			$Galaxy$ & $Type$ & $L_{[3.6]}$ & $R_{eff}$ & $R_{disk}$ & $M_{[gas]}$ & $B-V$ & $M_*/L_{[3.6]}$ & $M_{vir}$ & $c$ & $\chi^2_{\nu}$ \\
			& & ($10^9L_\odot$)& $(kpc)$& $(kpc)$& ($10^9M_\odot$)& & & ($10^{10}M_\odot$)& &  \\ \hline
			
			UGC12506 & 6 & 139.571$\pm$ 3.214 & 12.36 & 7.38 & 35.56 &  & $0.29_{-0.25}^{+0.27}$ & $12.41_{-1.67}^{+1.8}$ & $21.45_{-1.10}^{+1.07}$ & 0.15 \\
			UGC12632 & 9 & 1.301$\pm$ 0.030 & 3.94 & 2.42 & 1.74 & 0.68 & $2.37_{-0.48}^{+0.50}$ & $80.35_{-23.06}^{+27.91}$ & $3.85_{-0.36}^{+0.33}$ & 0.25 \\
			UGC12732 & 9 & 1.667$\pm$ 0.048 & 3.12 & 1.98 & 3.66 &  & $1.66_{-0.50}^{+0.53}$ & $459_{-92}^{+108}$ & $3.08_{-0.18}^{+0.18}$ & 0.13 \\
			UGCA281 & 11 & 0.194$\pm$ 0.007 & 1.57 & 1.72 & 0.06 &  & $0.87_{-0.23}^{+0.24}$ & $2971_{-1122}^{+1509}$ & $1.86_{-0.20}^{+0.19}$ & 0.39 \\
			UGCA442 & 9 & 0.140$\pm$ 0.005 & 1.71 & 1.18 & 0.26 &  & $1.38_{-0.84}^{+0.87}$ & $210_{-20}^{+21}$ & $3.47_{-0.09}^{+0.09}$ & 2.40 \\ \hline
			
		\end{tabular}
		\caption{Continued}
	\end{center}
	
\end{table*}

\begin{table}
\caption{ Predicted $c - M_{vir}$ relation (Eq. \ref{c-m}) from the present analysis and other works. Column 2 and 3 shows the best-fitting coefficient for different  analysis (see text for more details). The last two column gives the corresponding reduced intrinsic scatter (SE) and P-values.}
\begin{tabular}{cccccccc}
  \hline
  Name                  & $\alpha$             &   $ \beta $              &  $ \chi^2 $ &   P-value    \\
  \hline
  \hline
   Bullock et al (2002) &   $-0.13$              &     1.13                  &        0.40    &    $4.97 \times 10^{-12}$           \\
   Duffy et al (2007)   &    $-0.10$              &    0.76                  &        0.41    &    $1.39 \times 10^{-14}$           \\
   Dutton et al (2014)   &   $-0.10$             &     1.12                   &      0.45   &    $8.88 \times 10^{-16}$          \\
   Sereno et al. (2015)   &  $-0.59$   &               2.50                    &   1.22         &    $ 2.20 \times 10^{-16} $\\
   This work            & $-0.30 $    &    $1.1 $      &      0.20   &         0.43                       \\
\hline
\label{t2}
\end{tabular}
\end{table}

\begin{figure*}
\includegraphics[width=180mm]{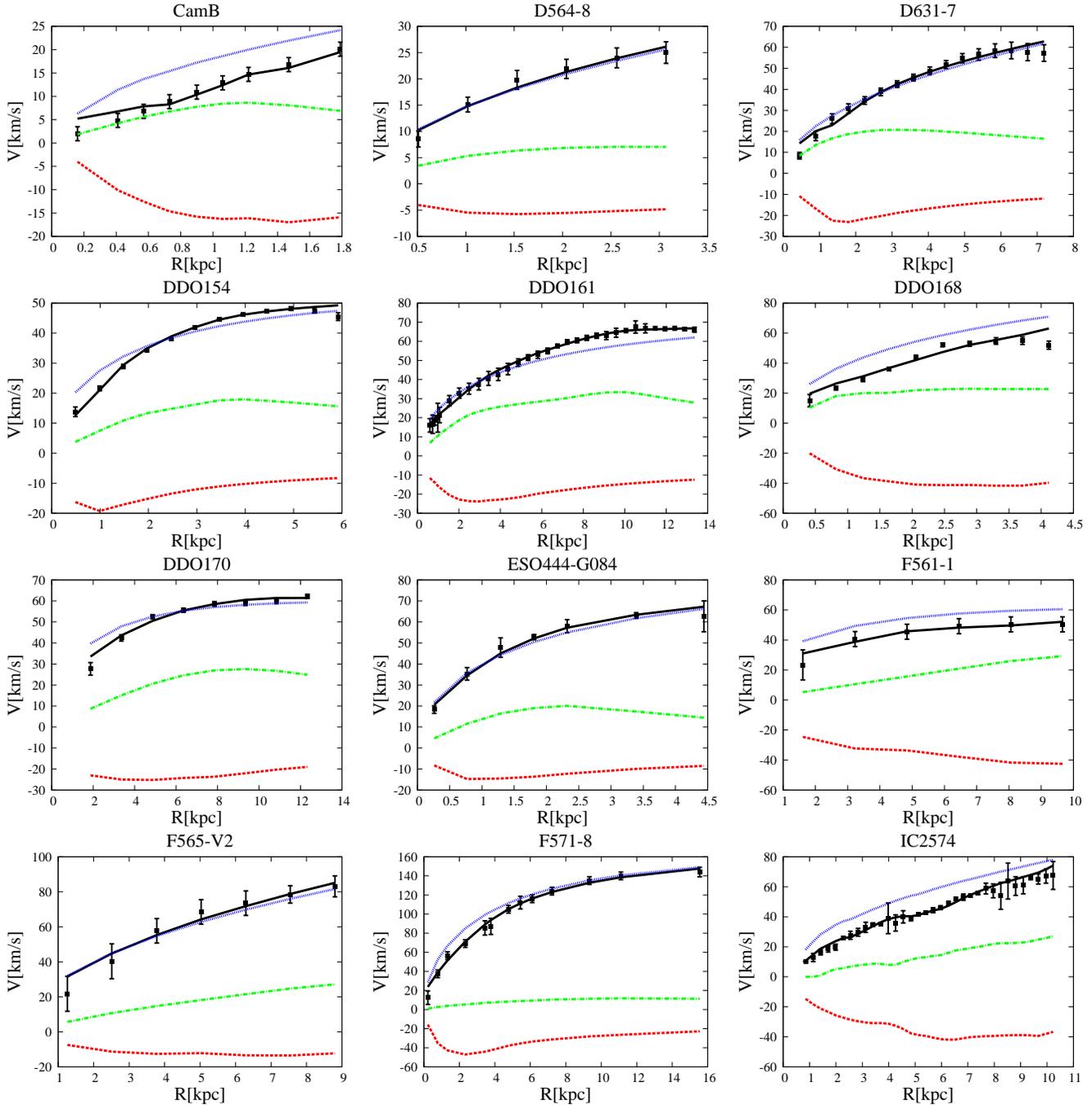}
\caption{   Three-parameter dark-matter halo fits (black curves) to the rotation curves of a subset of our sample of galaxies listed in Table \ref{t1}. Only cases with the negative $M_*/L_{[3.6]}$ ratios are shown. The data points represent the measured rotational velocities and their errors. The rotation curves of the individual components are also shown. The green curves show the contributions of HI gas to the rotation curves. The blue curves give the contribution of the dark halo. The contribution of the stellar disc and bulge (if any) to the best-fitting negative $M_*/L_{[3.6]}$ values is not shown. The fitting parameters are the mass-to-light ratio of the disk ($M_*/L_{[3.6]}$), the concentration (c), and the virial mass ($M_{vir}$) of the NFW halo. The solid black lines give the best fit with both $M_*/L_{[3.6]}$ ratio and halo as free parameters. The details of best-fitting parameters are listed in Table \ref{t1}. } \label{f2}
\end{figure*}

\begin{figure*}
\ContinuedFloat
\includegraphics[width=180mm]{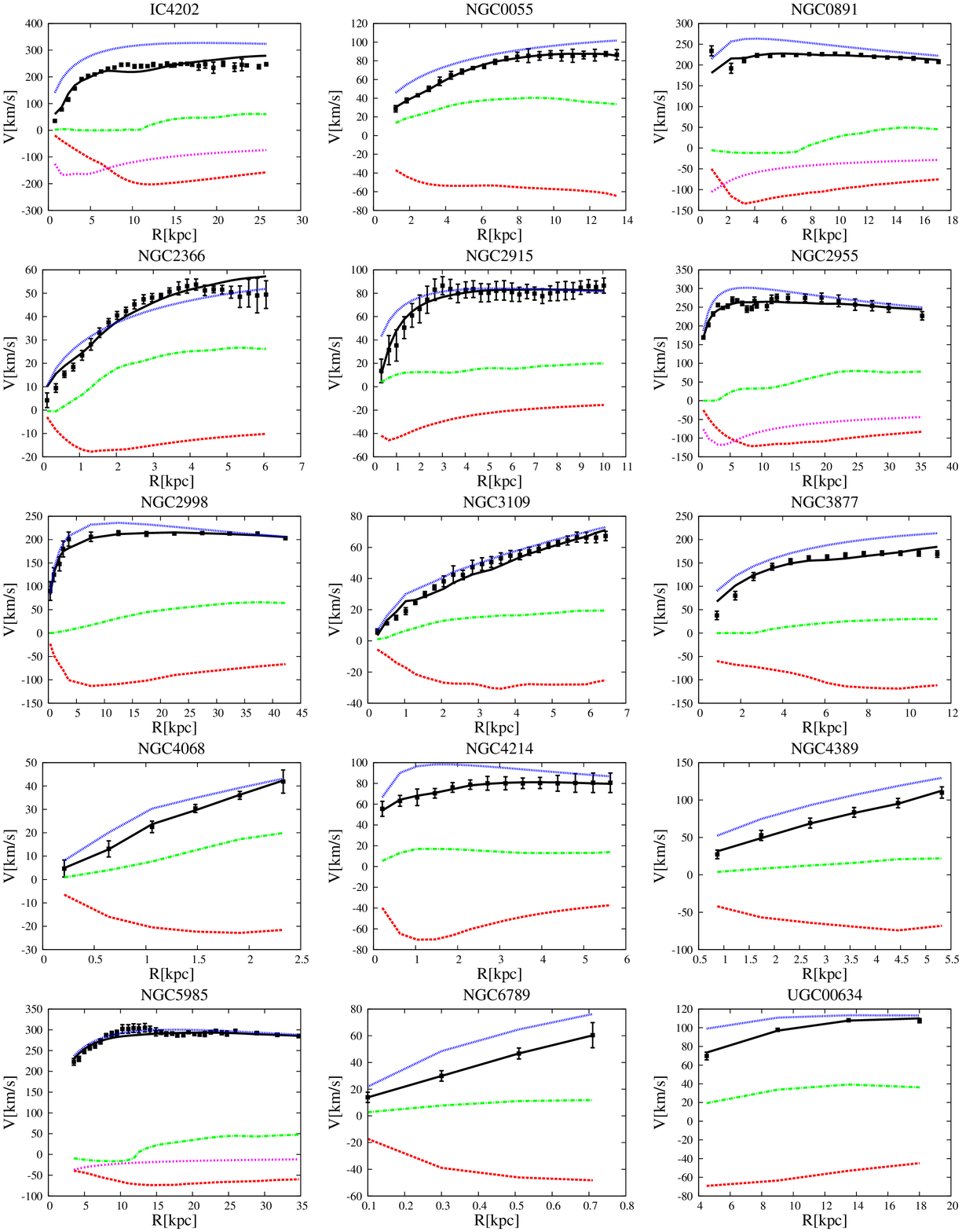}
\caption{Continued.
} \label{f3}
\end{figure*}

\begin{figure*}
\ContinuedFloat
\includegraphics[width=180mm]{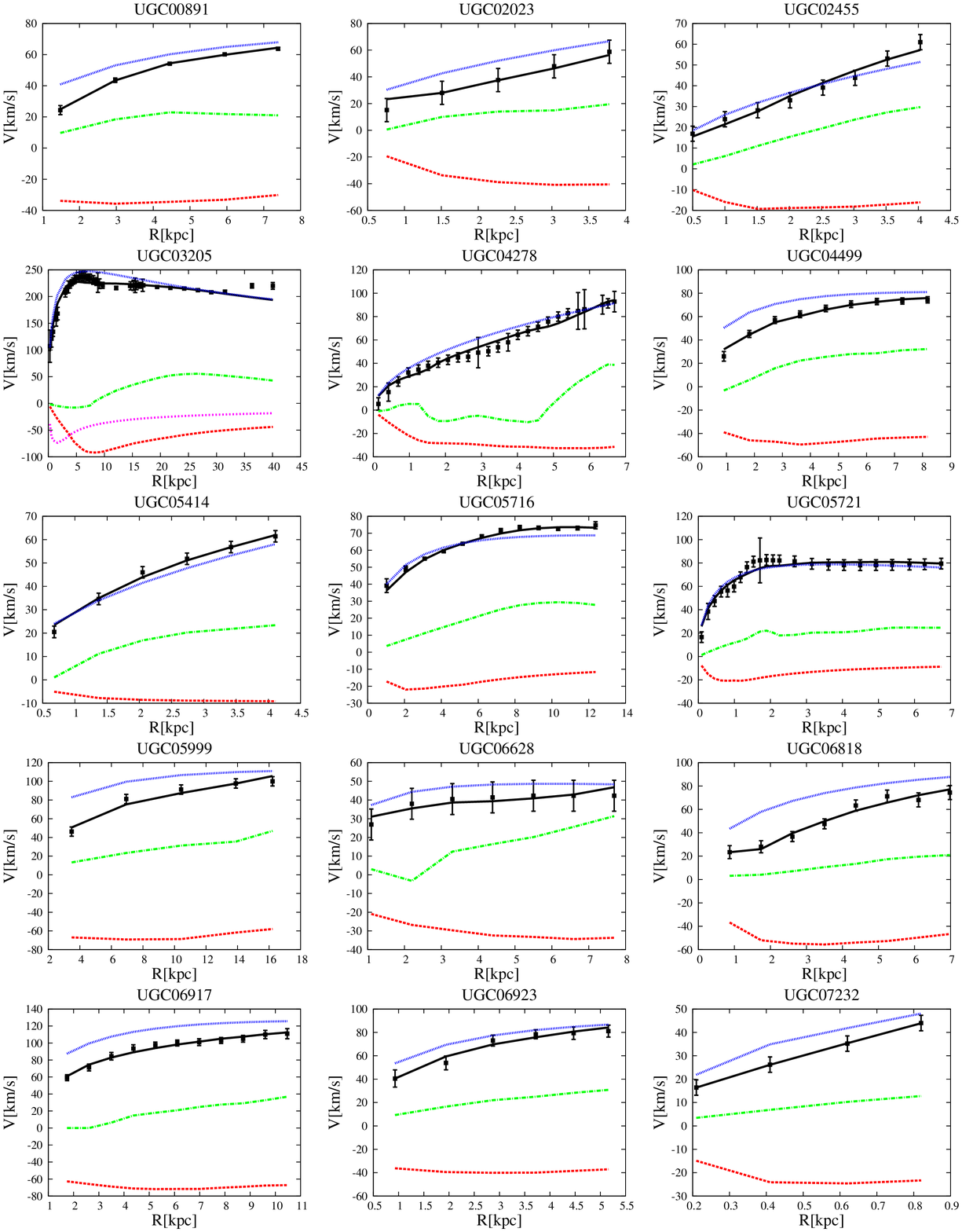}
\caption{Continued.
} \label{f4}
\end{figure*}

\begin{figure*}
\ContinuedFloat
\includegraphics[width=180mm]{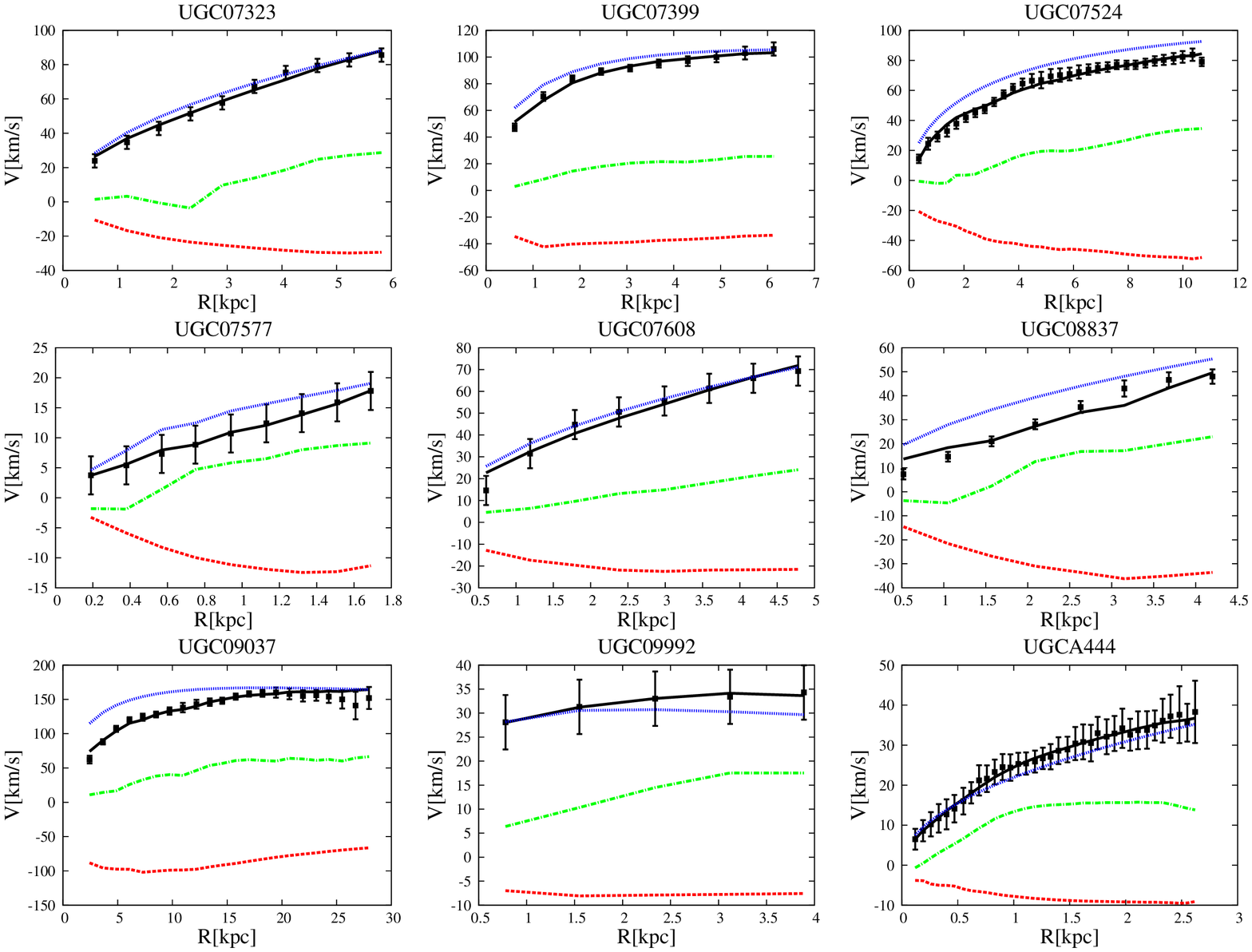}
\caption{Continued.
} \label{f5}
\end{figure*}

\section*{Acknowledgements}
PK acknowledges an ESO visiting-scientist grant during September and October 2017.

\bibliographystyle{plainnat}

\begin{thebibliography}{299}


\bibitem[\protect\citeauthoryear{Angus, Diaferio, \& Kroupa}{2011}]{Angus11} Angus, G.~W., Diaferio, A., \& Kroupa, P.\ 2011, MNRAS, 416, 1401

\bibitem[\protect\citeauthoryear{Begeman, Broeils, \& Sanders}{1991}]{beg91} Begeman K.~G., Broeils A.~H., Sanders R.~H., 1991, MNRAS, 249, 523

\bibitem[\protect\citeauthoryear{Bell \& de Jong}{2001}]{bel01} Bell E.~F., de Jong R.~S., 2001, ApJ, 550, 212
\bibitem[\protect\citeauthoryear{Bell et al}{2003}]{bell03} Bell, E. F.,McIntosh, D. H., Katz, N., \& Weinberg,M. D. 2003, ApJS, 149, 289
\bibitem[\protect\citeauthoryear{Bosma}{1978}]{bos78} Bosma A., 1978, PhD thesis, PhD Thesis, Groningen Univ.

\bibitem[\protect\citeauthoryear{Bottema}{1997}]{bott97} Bottema, R. 1997, A\&A, 328, 517

\bibitem[\protect\citeauthoryear{Bottema et al.}{2002}]{bot02} Bottema R., Pesta{\~n}a J.~L.~G., Rothberg B., Sanders R.~H., 2002, A\&A, 393, 453

\bibitem[\protect\citeauthoryear{Chemin, de Blok, \& Mamon}{2011}]{chem11} Chemin L., de Blok W.~J.~G., Mamon G.~A., 2011, AJ, 142, 109

\bibitem[\protect\citeauthoryear{Concas et al.}{2017}]{Concas17} Concas, A., Popesso, P., Brusa, M., et al. 2017, aap, 606, A36

\bibitem[\protect\citeauthoryear{Bertone et al}{2005}]{bert05} 	
	Bertone, G.; Hooper, D.; Silk, J., 2005,  Physics Reports, Volume 405, Issue 5-6, p. 279-390

\bibitem[\protect\citeauthoryear{Bullock et al}{2001}]{bull01} Bullock, J. S., Kolatt, T. S., Rachel, Y. S., et al. 2001, MNRAS, 321, 559

\bibitem[\protect\citeauthoryear{Kormendy, J. and Knapp, G. R}{1987}]{korknap87} Kormendy, J. and Knapp, G. R, 1987, Science, Vol.236, NO. 4806, 1360,
	
\bibitem[\protect\citeauthoryear{de Blok, McGaugh, \& Rubin}{2001}]{deb01} de Blok E., McGaugh S., Rubin V., 2001, astro, arXiv:astro-ph/0107366

\bibitem[\protect\citeauthoryear{de Blok \& Bosma}{2002}]{deb02} de Blok W.~J.~G., Bosma A., 2002, A\&A, 385, 816

\bibitem[\protect\citeauthoryear{de Blok et al.}{2008}]{deb08} de Blok W.~J.~G., Walter F., Brinks E.,
Trachternach C., Oh S.-H., Kennicutt R.~C., Jr., 2008, AJ, 136, 2648


\bibitem[\protect\citeauthoryear{Di Cintio et al}{2014}]{dici14} Di Cintio, A., Brook, C. B., Maccio, A. V., et al. 2014, MNRAS, 437, 415

\bibitem[\protect\citeauthoryear{Duffy et al}{2008}]{duff08} Duffy, A. R., Schaye, J., Kay, S. T., \& Dalla Vecchia, C. 2008, MNRAS, 390, L64

\bibitem[\protect\citeauthoryear{Dutton et al}{2014}]{dutt14} Dutton, A. A. \& Maccio, A. V. 2014, MNRAS, 441, 3359



\bibitem[\protect\citeauthoryear{Gentile et al.}{2004}]{gen04} Gentile G., Salucci P., Klein U., Vergani
D., Kalberla P., 2004, MNRAS, 351, 903

\bibitem[\protect\citeauthoryear{Gentile et al.}{2005}]{gen05} Gentile G., Burkert A., Salucci P., Klein
U., Walter F., 2005, ApJ, 634, L145

\bibitem[\protect\citeauthoryear{Gentile et al.}{2007}]{gen07a} Gentile G., Famaey B., Combes F., Kroupa P., Zhao H.~S., Tiret O., 2007, A\&A, 472, L25

\bibitem[\protect\citeauthoryear{Gentile, Tonini, \& Salucci}{2007}]{gen07b} Gentile G., Tonini C., Salucci P., 2007, A\&A, 467, 925

\bibitem[\protect\citeauthoryear{Gentile et al.}{2007}]{gen07c} Gentile G., Salucci P., Klein U., Granato
G.~L., 2007, MNRAS, 375, 199

\bibitem[\protect\citeauthoryear{Gnedin \& Zhao}{2002}]{Gnedin02} Gnedin O. Y., Zhao H., 2002, MNRAS, 333, 299

\bibitem[\protect\citeauthoryear{Ibata et al.}{2014}]{Ibata14}
Ibata N.~G., Ibata R.~A., Famaey B., Lewis G.~F., 2014, Natur, 511, 563


\bibitem[\protect\citeauthoryear{Katz et al}{2016}]{katz16} Katz, H., Lelli, F., McGaugh, S. S., et al. 2016,MNRAS, 466, 1648

\bibitem[\protect\citeauthoryear{Klypin et al}{2011}]{klypin11} Klypin, A. A., Trujillo-Gomez, S., \& Primack, J. 2011, ApJ, 740, 102

\bibitem[\protect\citeauthoryear{Kroupa}{2001}]{Kroupa01} Kroupa P., 2001, MNRAS, 322, 231

\bibitem[\protect\citeauthoryear{Kroupa et al.}{2010}]{kro10} Kroupa P., et al., 2010, A\&A, 523, A32

\bibitem[\protect\citeauthoryear{Kroupa et al.}{2013}]{kro13} Kroupa, P., Weidner, C., Pflamm-Altenburg, J., Thies I.,
Dabringhausen J., Marks M., Maschberger T., 2013, Planets, Stars and Stellar Systems. Volume 5: Galactic Structure and Stellar Populations, 115 (astro-ph/1112.3340)

\bibitem[\protect\citeauthoryear{Kroupa }{2015}]{kro15} Kroupa P., 2015, CaJPh, 93, 169

\bibitem[\protect\citeauthoryear{Lelli et al. }{2014}]{Lelli14} Lelli, F., Verheijen, M., \& Fraternali, F.\ 2014, aap, 566, A71

\bibitem[\protect\citeauthoryear{Lelli, McGaugh, \& Schombert}{2016}]{Lelli16}
Lelli F., McGaugh S.~S., Schombert J.~M., 2016, AJ, 152, 157


\bibitem[\protect\citeauthoryear{Lovell et al.}{2018}]{Lovell18} Lovell, M.~R., Pillepich, A., Genel, S., et al. 2018, arXiv:1801.10170

\bibitem[\protect\citeauthoryear{L{\"u}ghausen et al. }{2015}]{lug15}
L{\"u}ghausen, F., Famaey, B., \& Kroupa, P.\ 2015, Canadian Journal of Physics, 93, 232

\bibitem[\protect\citeauthoryear{Maccio et al}{2008}]{macc08} Maccio, A. V., Dutton, A. A., \& van den Bosch, F. C. 2008, MNRAS, 391, 1940
\bibitem[\protect\citeauthoryear{McGaugh \& de Blok}{1998}]{mcg98} McGaugh S.~S., de Blok W.~J.~G., 1998, ApJ, 499, 41

\bibitem[\protect\citeauthoryear{McGaugh et al.}{2007}]{mcg07}
McGaugh S.~S., de Blok W.~J.~G., Schombert J.~M., Kuzio de Naray R., Kim J.~H., 2007, ApJ, 659, 149

\bibitem[\protect\citeauthoryear{McGaugh}{2016}]{McGaugh16}
McGaugh, S.~S.\ 2016, ApJ, 816, 42


\bibitem[\protect\citeauthoryear{Moore et al.}{1999}]{moo99} Moore B., Ghigna S., Governato F., Lake G., Quinn T., Stadel J., Tozzi P., 1999, ApJ, 524, L19

\bibitem[\protect\citeauthoryear{Navarro, Frenk, \& White}{1996}]{nav96} Navarro J.~F., Frenk C.~S., White S.~D.~M., 1996, ApJ, 462, 563

\bibitem[\protect\citeauthoryear{Navarro et al.}{2004}]{nav04} Navarro J.~F., et al., 2004, MNRAS, 349, 1039

\bibitem[\protect\citeauthoryear{Neto et al.}{2007}]{neto07} Neto, A. F., Gao, L., Bett, P., et al. 2007, MNRAS, 381, 1450

\bibitem[\protect\citeauthoryear{Oehm et al.}{2017}]{Oehm17} Oehm, W., Thies, I., \& Kroupa, P. 2017, mnras, 467, 273

\bibitem[\protect\citeauthoryear{Oh et al.}{2008}]{Oh08} Oh
S.-H., de Blok W.~J.~G., Walter F., Brinks E., Kennicutt R.~C., Jr., 2008,
AJ, 136, 2761

\bibitem[\protect\citeauthoryear{Pace}{2016}]{pace16} Pace, A. B. 2016, ArXiv e-prints

\bibitem[\protect\citeauthoryear{Portinari, Sommer-Larsen, \& Tantalo}{2004}]{por04} Portinari L., Sommer-Larsen J., Tantalo R., 2004, MNRAS, 347, 691

\bibitem[\protect\citeauthoryear{Pawlowski, Pflamm-Altenburg,
\& Kroupa}{2012}]{Pawlowski12} Pawlowski M.~S., Pflamm-Altenburg J., Kroupa P., 2012, MNRAS, 423, 1109

\bibitem[\protect\citeauthoryear{Persic, Salucci,
\& Stel}{1996}]{per96} Persic M., Salucci P., Stel F., 1996, MNRAS, 281, 27

\bibitem[\protect\citeauthoryear{Randriamampandry \& Carignan}{2014}]{carignan14} Randriamampandry T.~H., Carignan C., 2014, MNRAS, 439, 2132


\bibitem[\protect\citeauthoryear{Read \& Gilmore}{2005}]{Read05} Read J. I., Gilmore G., 2005, MNRAS, 356, 107

\bibitem[\protect\citeauthoryear{Rubin, Ford,
\& D'Odorico}{1970}]{rub70} Rubin V.~C., Ford W.~K., Jr., D'Odorico S., 1970, ApJ, 160, 801

\bibitem[\protect\citeauthoryear{Rubin, Thonnard \& Ford}{1978}]{rub78}
Rubin V. C., Thonnard N., Ford J. W. K., 1978, ApJ, 225, L107

\bibitem[Schombert \& McGaugh(2014)]{Schombert2012} Schombert, J., \& McGaugh, S.\ 2014, pasa, 31, e036

\bibitem[\protect\citeauthoryear{Sereno et al}{2015}]{ser2015} Sereno, M., Giocoli, C., Ettori, S., Moscardini, L., 2015, MNRAS, 449, 2024

\bibitem[\protect\citeauthoryear{Wechsler et al}{2002}]{wechsler02} Wechsler, R. H., Bullock, J. S., Primack, J. R., Kravtsov, A. V., \& Dekel, A. 2002, ApJ, 568, 52

\bibitem[\protect\citeauthoryear{Wu \& Kroupa}{2013}]{wu13} Wu X., Kroupa P., 2015, MNRAS, 435, 728,

\bibitem[\protect\citeauthoryear{Wu \& Kroupa}{2015}]{wu15} Wu X., Kroupa P., 2015, MNRAS, 446, 330


\bibitem[\protect\citeauthoryear{Zonoozi \& Haghi}{2010}]{zon10} Zonoozi A.~H., Haghi H., 2010, A\&A, 524, A53
%
\bibitem[\protect\citeauthoryear{Zwicky}{1933}]{Zwic} Zwicky F., 1933, AcHPh, 6, 110
%

\end{thebibliography}

\bsp \label{lastpage} \end{document}